\documentclass[aps,prd,tightenlines,superscriptaddress,nofootinbib,twocolumn]{revtex4}

\usepackage[utf8]{inputenc}
\usepackage{tikz}
\usepackage{graphicx}
\usepackage{subfigure}
\usepackage{amsmath}
\usepackage{amssymb}
\usepackage{color}
\usepackage{appendix}
\usepackage[colorlinks=true]{hyperref}
\usepackage[acronym,shortcuts]{glossaries}
\usepackage{placeins}
\usepackage{url}

\newcommand{\sub}[1]{\ensuremath{_{\textnormal{#1}}}}

\newacronym{BDT}{BDT}{boosted decision tree}
\newacronym{BNS}{BNS}{binary neutron star inspiral}
\newacronym{CSG}{CSG}{circular sine-Gaussian}
\newacronym{chirplet}{chirplet}{chirplet}
\newacronym{FAP}{FAP}{false alarm probability}
\newacronym{GRB}{GRB}{gamma ray burst}
\newacronym{GW}{GW}{gravitational-wave}
\newacronym{GWB}{GWB}{gravitational-wave burst}
\newacronym{MVA}{MVA}{multivariate analysis}
\newacronym{NN}{NN}{neural network}
\newacronym{TMVA}{TMVA}{toolkit for multivariate analysis}
\newacronym{WNB}{WNB}{White noise burst}
\glsdisablehyper

\begin{document}

\title{Gravitational-Wave Detection using Multivariate Analysis}

\author{Thomas~S.~Adams}
\affiliation{School of Physics and Astronomy, Cardiff University, Cardiff, United Kingdom, CF24 3AA}
\author{Duncan~Meacher}
\affiliation{ARTEMIS, UMR7250, Université de Nice Sophia Antipolis, CNRS, Observatoire de la Côte d’Azur, 06300, Nice, France}
\affiliation{School of Physics and Astronomy, Cardiff University, Cardiff, United Kingdom, CF24 3AA}
\author{James~Clark}
\affiliation{University of Massachusetts - Amherst, Amherst, MA 01003, USA}
\affiliation{School of Physics and Astronomy, Cardiff University, Cardiff, United Kingdom, CF24 3AA}
\author{Patrick~J.~Sutton}
\affiliation{School of Physics and Astronomy, Cardiff University, Cardiff, United Kingdom, CF24 3AA}
\author{Gareth~Jones}
\affiliation{School of Physics and Astronomy, Cardiff University, Cardiff, United Kingdom, CF24 3AA}
\author{Ariana~Minot}
\affiliation{Harvard University, Cambridge, MA 02138, USA}

\begin{abstract}
Searches for gravitational-wave bursts (transient signals, typically of unknown waveform) require identification of weak signals in background detector noise.
The sensitivity of such searches is often critically limited by non-Gaussian noise fluctuations which are difficult to distinguish from real signals, posing a key problem for transient gravitational-wave astronomy.
Current noise rejection tests are based on the analysis of a relatively small number of measured properties of the candidate signal, typically correlations between detectors.
\Ac{MVA} techniques probe the full space of measured properties of events in an attempt to maximise the power to accurately classify events as signal or background.
This is done by taking samples of known background events and (simulated) signal events to train the \ac{MVA} classifier, which can then be applied to classify events of unknown type.
We apply the \ac{BDT} \ac{MVA} technique to the problem of detecting gravitational-wave bursts associated with gamma-ray bursts.
We find that \acp{BDT} are able to increase the sensitive distance reach of the search by as much as 50\%, corresponding to a factor of $\sim3$ increase in sensitive volume.
This improvement is robust against trigger sky position, large sky localisation error, poor data quality, and the simulated signal waveforms that are used.
Critically, we find that the \ac{BDT} analysis is able to detect signals that have different morphologies to those used in the classifier training and that this improvement extends to false alarm probabilities beyond the 3$\sigma$ significance level.
These findings indicate that \ac{MVA} techniques may be used for the robust detection of gravitational-wave bursts with {\em a priori} unknown waveform.
\end{abstract}

\maketitle

\section{Introduction}\label{sec:intro}

The upcoming Advanced LIGO \cite{Harry:2010zz}, Advanced Virgo \cite{TheVirgoCollaboration:ug}, and KAGRA \cite{Somiya:2012en} gravitational-wave detectors will open a new channel for studying the most extreme phenomena and environments found in nature, including \acp{GRB} \cite{2006RPPh...69.2259M}, core-collapse supernovae \cite{Ott09,Kotake:2011vg}, and neutron stars \cite{2009ApJ...702.1171C}.
The associated \ac{GW} emission typically depends on poorly understood physics, such as the equation-of-state of matter at supra-nuclear densities.
While \acp{GW} will therefore provide an exciting new probe of these astrophysical systems, the detection of a \ac{GW} burst depends on being able to distinguish a rare weak signal with {\em a priori} unknown waveform from the highly non-stationary and non-Gaussian background noise of the detectors.

Unfortunately, in the analysis of data from the first-generation LIGO and Virgo detectors no \ac{GW} signals were detected; the only case where a (simulated) \ac{GW} signal of realistic amplitude was identified at a significance level permitting a tentative detection claim ($>3\sigma$) relied on the precise knowledge of the complex time-frequency structure of the signal (a binary neutron star merger) to reject spurious background transients \cite{lowMassS5y2}\cite{lowMassS6}.
Similar blind injection tests of general burst searches, where the \ac{GW} emission is not known {\em a priori} have shown that current model-independent methods are not able to reduce the false alarm rate significantly below 0.1-1 per year \cite{burstS5y2,burstS6allsky} This points to a clear need to investigate new techniques for signal/background discrimination.

In \ac{GW} transient searches, candidate events (clusters of excess power in the signal streams) are typically classified as signal or background by applying thresholds or rankings based on a small number of the measured properties of the events, such as signal-to-noise ratio, $\chi^{2}$ match to a model \cite{Allen05}, or cross-correlation between detectors \cite{Was:2012ey}.
By contrast \acf{MVA} methods \cite{Hoecker:2007tp,Hastie2001,webb2002} based on machine learning techniques explore the full dimensionality of the event space, and have proven useful in fields of research where there is a need to separate signal from background in large quantities of high-dimensional data, such as particle physics \cite{Collaboration:2012vb,Wolter:ta}.
Also, \ac{MVA} methods have previously been shown to give improvements in detection probability, for a given false alarm rate, when applied to a \ac{GW} search \cite{0264-9381-25-10-105024}.

\Ac{MVA} addresses the problem of signal/background classification through supervised machine learning.
Starting with samples of data of known type (signal or background), the data are divided randomly into {\em training} and {\em testing} sets, each of which consists of a mix of signal and background events.
The training data set is used to fix the parameters of the \ac{MVA} classifier, a function that assigns to each input event a measure of its consistency with the signal or background hypotheses.
The parameters are chosen to achieve the best separation between the signal and background training events.
The trained classifier is then applied to the testing data set to obtain an unbiased evaluation of the classifier's detection performance from the number of correctly classified signal and background events.

In this paper we investigate the power of \ac{MVA} to detect \ac{GW} bursts.
Specifically, we apply the \ac{TMVA} package \cite{Hoecker:2007tp} to events from the analysis of LIGO and Virgo data associated with \acp{GRB}.
By reclassifying events with the \acf{BDT} \ac{MVA} technique, we find that \acp{BDT} are able to suppress background events relative to signal events and increase the sensitive distance reach of the search by as much as 50\%.
We see consistent improvement, regardless of the sky position and position uncertainty of the \ac{GRB} and good/poor data quality, and for a variety of signal morphologies.
Critically, we find that the \ac{BDT} analysis is able to detect signals that have different morphologies to those used in the classifier training; in the worst case, a classifier trained with the ``wrong'' signal morphology is as sensitive as the standard LIGO--Virgo analysis.
A detailed study of one event shows that the suppression of the background by \acp{BDT} extends at least down to false alarm probabilities of order $10^{-5}$ for a single \ac{GRB}.
This corresponds to a $3\sigma$ significance or better in the context of a LIGO--Virgo search, which typically analyses 100--150 \acp{GRB}.
%
These results indicate that \ac{MVA} may be a promising technique for robust \ac{GW} burst detection.

This paper is organised as follows.
In Section~\ref{sec:xpipeline} we briefly review the standard \ac{GW} transient analysis package, \textsc{X-Pipeline}.
In Section~\ref{sec:multivariate_analysis} we describe \ac{MVA} techniques, focussing on the \ac{BDT} classifier which we use in this paper.
In Section~\ref{sec:signalpopulation} we describe the waveforms used for our signal event populations.
In Section~\ref{sec:results} we describe the various scenarios used for our tests and compare the performance of \ac{BDT} and \textsc{X-Pipeline}.
We discuss the results and summarise our conclusions in Section~\ref{sec:summary}.

\section{\textsc{X-Pipeline}}
\label{sec:xpipeline}

\textsc{X-Pipeline} is a standard analysis package used for LIGO--Virgo searches for generic \ac{GW} transients associated with \acp{GRB} and other astrophysical triggers.
Here we give a brief description of the aspects of the pipeline relevant for this paper; for a complete description see \cite{Sutton10,Was:2012ey}

\textsc{X-Pipeline} processes data from a network of \ac{GW} detectors.
First, the data are time-shifted according to the direction of the \ac{GRB} trigger so that \ac{GW} signals will arrive simultaneously in all data streams.
Various combinations of the data streams are then formed, split into two groups: those that maximise the signal-to-noise ratio of a \ac{GW} (signal streams); and those that cancel out \ac{GW} signals leaving only noise events (null streams).
Time-frequency maps of the signal streams are constructed, and clusters of pixels that have large energy values are selected as candidate signal events \cite{Sutton10}.
For each event cluster a variety of energy measures and time-frequency information (such as peak frequency, bandwidth, peak time, duration, and number of pixels) are recorded.
Each event is also assigned a significance measure based on the energy in the signal stream; in this study we use a Bayesian-inspired likelihood statistic appropriate for circularly polarised \acp{GW} \cite{Was:2012ey}.
Brief descriptions of the 15 event properties fed into the \ac{BDT} analysis are given in Table.~\ref{tab:properties}.

\begin{table}
\begin{center}
\begin{tabular}{l|l}
    \hline
    \hline
    Event property & Description\\
    \hline
    \hline
    significance    & The statistic used to rank events. By \\
                    & default equal to loghbayesiancirc.\\
    \hline
    loghbayesiancirc & Bayesian-inspired likelihood ratio for the\\
                    & hypothesis of a circularly polarised \acp{GW}\\
                    & versus Gaussian noise \cite{Was:2012ey}.\\
    \hline
    $E\sub{max}$    & The maximum amount of energy in the\\
                    & whitened data that is consistent with the\\
                    & hypothesis of a \ac{GW} of any polarisation\\
                    & from a given sky position.\\
    \hline
    $E\sub{circ}$   & The circular coherent energy is the\\
                    & maximum amount of energy in the\\
                    & whitened data that is consistent with the\\
                    & hypothesis of a circularly polarised\\
                    & \ac{GW} from a given sky position.\\
    \hline
    $I\sub{circ}$   & The circular incoherent energy is the\\
                    & sum of the autocorrelation terms of $E\sub{circ}$;\\
                    & i.e., neglecting cross-correlation terms.\\
    \hline
    $E\sub{circnull}$ & The circular coherent null energy,\\
                      & $E\sub{max}-E\sub{circ}$. Physically, it is the energy\\
                      & in the whitened data that is inconsistent\\
                      & with the hypothesis of a circularly polarised\\
                      & \ac{GW} from a given sky position, but which\\
                      & could be produced by a \ac{GW} of a\\
                      & different polarisation.\\
    \hline
    $I\sub{circnull}$ & The circular incoherent null energy is the\\
                    & sum of the autocorrelation terms of\\
                    & $E\sub{circnull}$; i.e., neglecting cross-correlation\\
                    & terms.\\
    \hline
    $E\sub{null}$   & The coherent null energy is the minimum\\
                    & amount of energy in the whitened data\\
                    & that is inconsistent with the hypothesis\\
                    & of a \ac{GW} of any polarisation from a given\\
                    & sky position.\\
    \hline
    $I\sub{null}$   & The incoherent null energy is the\\
                    & sum of the autocorrelation terms of $E\sub{null}$;\\
                    & i.e., neglecting cross-correlation terms.\\
    \hline
    $E\sub{H1}$      & The cluster energy in the LIGO-Hanford\\
                    & interferometer.\\
    \hline
    $E\sub{L1}$      & The cluster energy in the LIGO-Livingston\\
                    & interferometer.\\
    \hline
    $E\sub{V1}$      & The cluster energy in the Virgo\\
                    & interferometer.\\
    \hline
    number of pixels & The number of pixels in the cluster.\\
    \hline
    duration         &  The extent of the cluster in time (s).\\
    \hline
    bandwidth        & The extent of the cluster in frequency (Hz).\\
    \hline
    \hline
\end{tabular}
\end{center}
\caption{\label{tab:properties}Cluster properties recorded by \textsc{X-Pipeline} that are fed into the \ac{BDT} analysis. See \cite{Sutton10,thesisWas,Was:2012ey} for more details.}
\end{table}

\begin{figure}
  \includegraphics[width=0.45\textwidth]{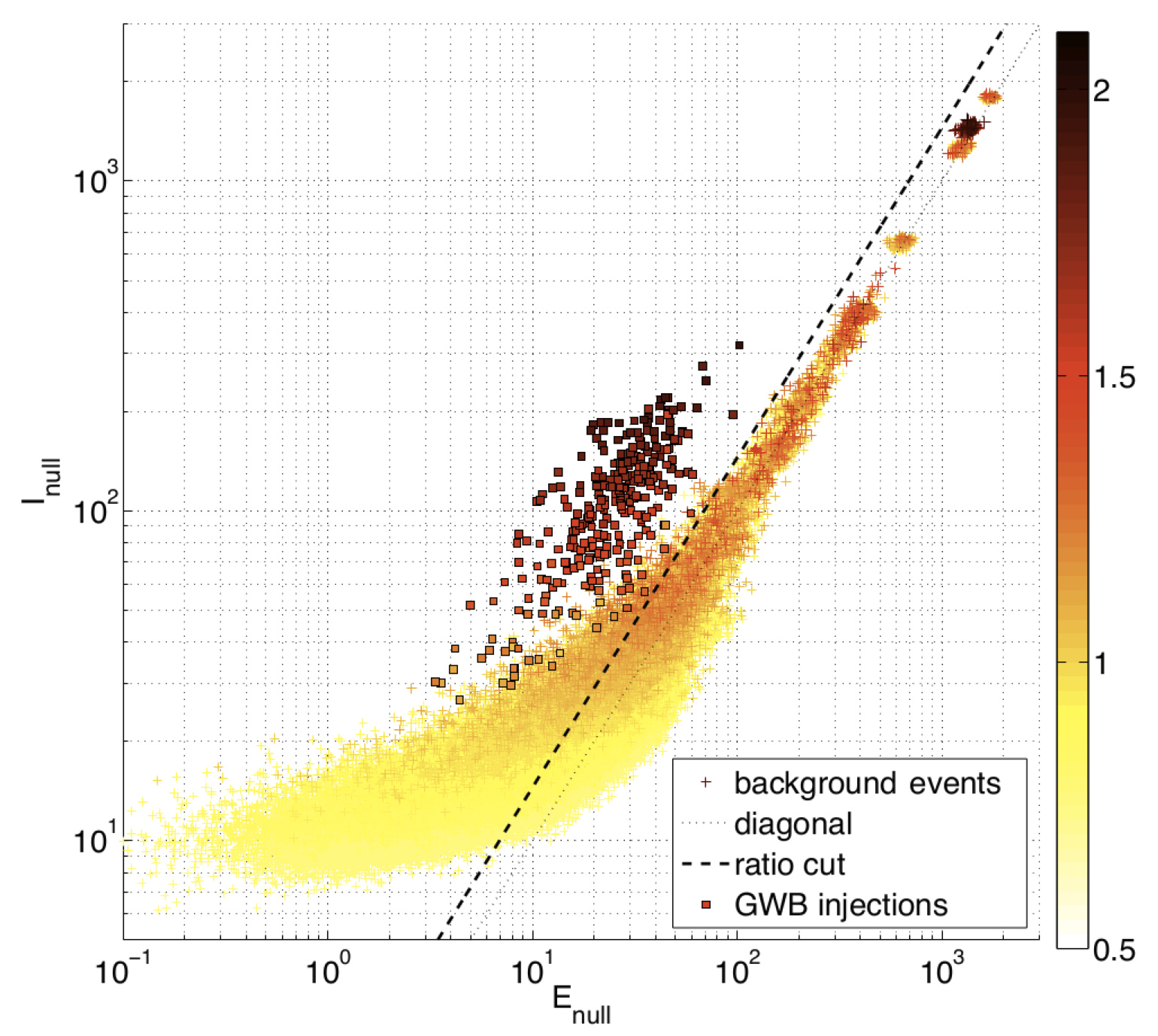}
  \caption{\label{fig:xpipeIvsE}$I\sub{null}$ vs. $E\sub{null}$ for background events ($+$) and simulated gravitational wave bursts ($\Box$) for a sample \ac{GRB} search.
  The colour of the symbols represents the significance associated with each event, with redder (darker) colours representing higher significance.
  The dotted line shows the $I=E$ diagonal.
  The dashed line shows the threshold selected by \textsc{X-Pipeline} to separate background and signal; all events that fall below this line are discarded \cite{Sutton10}.}
\end{figure}

Background noise fluctuations produce clusters of excess power in the signal streams.
For these noise ``glitches'' there is typically a strong correlation between the energy in the individual detector data streams (incoherent energy $I$) and the corresponding energy in the combined detector data steams (coherent energy $E$) \cite{Chatterji06}.
These incoherent and coherent energies are compared in order to remove events with properties similar to the background noise.
The test uses a threshold curve in the two-dimensional $(I,E)$ space,
such as that shown in Fig.~\ref{fig:xpipeIvsE}.
The test may be single-sided, vetoing all events on one side of the line,
or two-sided, vetoing events inside a band centred on the $I=E$ diagonal.
Two curve shapes are tested (see \cite{thesisWas} for a discussion):
\begin{eqnarray}
    \centering
    \frac{I}{E} & = & \mathrm{constant}\, , \label{eqn:ratiocut} \\
    \frac{|E-I|}{(E+I)^{0.8}} & = & \mathrm{constant}\, . \label{eqn:alphacut}
\end{eqnarray}
For the studies performed in this paper, there are usually three distinct $(I,E)$ energy pairs available for testing: one associated with the signal stream, and two associated with null streams.
Which pairs will be used for a given analysis and the thresholds to be used are determined by an automated tuning procedure.

The thresholds for the background rejection tests are selected to optimise the trade-off between glitch rejection and signal acceptance.
Samples of known background events are generated by analysing data with unphysically large ($>1\,$s) relative time shifts applied to the detector data streams.
Known signal events are generated by adding simulated \ac{GW} signals to the data, known as ``injections''.
The background and injection events are randomly divided into two equal sets, one that is used for training the pipeline and a second that is used for testing performance.
For each $(I,E)$ pair the background rejection test is applied to both the background and injection training samples using a range of trial thresholds.
The cumulative distribution of significance of background events surviving the cuts is computed.
We then determine the minimum injection amplitude at which 50\% of the injections both survive the cuts and have significance greater than a user-specified fraction of the background (e.g., greater than 99\% of the background, for a \ac{FAP} $p\le0.01$).
The optimum thresholds are then defined as those which yield the lowest minimum injection amplitude at the user-specified \ac{FAP} (i.e., which make the analysis sensitive to the weakest \ac{GW} signals at fixed \ac{FAP}).
Finally, unbiased estimates of the background distribution and detectable injection amplitudes are made by processing the training data set with our fixed optimal test thresholds.

\section{Multivariate analysis}\label{sec:multivariate_analysis}

The sensitivity of \ac{GW} transient searches is limited by the ability to distinguish between signals and background.
As described above, the standard \textsc{X-Pipeline} analysis uses a simple pass/fail cut in one or more two-dimensional parameter spaces.
These cuts only discriminate between signal and background using a few of the variables associated with each event, and ignore other information such as duration, bandwidth, and time-frequency volume.
\ac{MVA} techniques can mine the full parameter space of the events to better discriminate between signal and background.
Here we explore the efficacy of \ac{MVA} in \ac{GW} detection by using the \ac{BDT} classifier to re-evaluate the significance of events from an \textsc{X-Pipeline} analysis.
We find that the \ac{BDT} classification of events renders the $(I,E)$ test redundant, and that \ac{BDT} improves the amplitude sensitivity of the analysis
by up to 50\% in some cases.

\subsection{Toolkit for multivariate analysis}
\label{sec:tmva}

We use the \textsc{ROOT} \cite{ROOT:2006vu} based software package \ac{TMVA} \cite{Hoecker:2007tp} which was developed by the particle physics community.
\ac{TMVA} takes as input known signal and background events.
These events are split randomly into two sets, one for training the classifier and the other for testing its performance.
This split ensures that the testing produces an unbiased estimate of the classifier performance, since the event used for testing are independent of those used for training.

The results from the training of the classifier are stored in a ``weight'' file, that contains all the information needed to evaluate the classifier function for any input event and assign a \ac{MVA} significance value.
This significance is a measure of the likelihood of an event being a signal; events with high values of significance are more likely to be signals, and events with small values of significance are more likely to be background.
The \ac{TMVA} package provides many classifiers such as \acfp{BDT}, \acfp{NN} and projective likelihood \cite{Hoecker:2007tp}.
Initial tests found that a number of the classifiers produced similar results, and that by tuning the classifier parameters an improvement of $\sim10\%$ was possible; for simplicity we selected \ac{BDT} using the default parameters for in-depth testing as it exhibited the best performance in the shortest processing time.
However, a more in-depth study should be performed to optimise \ac{MVA} performance for \ac{GW} applications.

\subsection{Boosted Decision Trees}
\label{sec:bdt}

\begin{figure}
\includegraphics[width=0.45\textwidth]{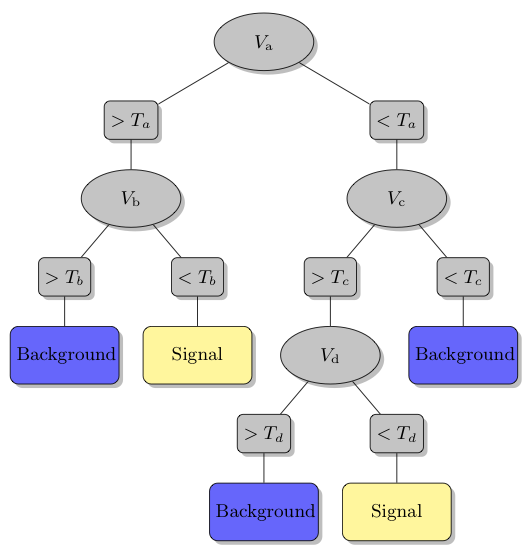}
\caption{\label{fig:decisiontree}Schematic of a decision tree.
Decision nodes for each event variable ($V\sub{a}$, $V\sub{b}$, $V\sub{c}$, and $V\sub{d}$) are grey ellipses, with thresholds ($T\sub{a}$, $T\sub{b}$, $T\sub{c}$, and $T\sub{d}$) for each branch given in grey rectangles.
The events are classified by a majority vote of events in a leaf node as either signal (yellow/ light grey rectangles) or background (blue /dark grey rectangles).}
\end{figure}

A decision tree consists of a series of yes/no decisions applied to each event, as shown schematically in Fig.~\ref{fig:decisiontree}.
Beginning at the root node, an initial criterion for splitting the full set of events is determined.
The split criterion consists of a threshold applied to a single variable, selected to best discriminate signal from background.
This split results in two branches, each containing a subset of the events.
The process then repeats, with a new split criterion being determined at each branch node to further separate signal from background.
The splitting process ends once a minimum number of events has been reached within a node, which then becomes a leaf node.
We use the default value in \ac{TMVA} of 400 events.
Leaf nodes are labelled as either signal or background depending on the class of the majority of training events that fall within it.
The user can specify criteria at which the tree stops being grown, such as how many layers a tree can contain and the total number of nodes which may be created.

Decision trees are susceptible to statistical fluctuations within the set of training events used to derive the tree structure.
To avoid over training, a whole ``forest'' of decision trees are created, each generated using a randomly selected subset of the training events.
The final classification of events is determined by a majority vote from the classifications of each individual tree within the forest.
This procedure stabilises the response of individual trees and enhances overall performance.
We use the default forest in \ac{TMVA} made of 400 \acp{BDT}.

Another procedure to statistically stabilise the classifier is ``boosting''.
During training, signal and background events which are misclassified by one tree are given increased weight when constructing the next tree in the forest.
We use the default boosting method in \ac{TMVA}, ``AdaBoost''.

\section{Signal population}
\label{sec:signalpopulation}

We test \ac{BDT} for a common \ac{GW} scenario: the search for a \ac{GW} burst associated with a \ac{GRB}.

\acp{GRB} are astrophysical events that are observed as an intense flash of gamma rays.
Short-duration ($< 2\,$s) \acp{GRB} are thought to be due to the merger of compact binaries consisting of two neutron stars or a neutron star and a black hole \cite{Duncan:1992hh,NAKAR:2007es}.
Long-duration ($> 2\,$s) \acp{GRB} are associated with the core collapse of massive stars \cite{Hjorth11}.
Both of these progenitor models are highly relativistic and lead to the formation of an accreting black hole (or possibly a magnetar \cite{2009ApJ...702.1171C}).
While the \acp{GW} produced by the inspiral phase of compact binaries coalescences in short \acp{GRB} are well modelled, the expected signal from long \acp{GRB} is speculative.
A number of such searches for \acp{GW} associated with \acp{GRB} have been performed by the LIGO and Virgo collaborations \cite{Abbott05,GRB070201,burstGrbS234,Acernese08,2010ApJ...715.1453A}, including several using \textsc{X-Pipeline} \cite{burstGrbS5,grb051103,Abadie:2012cf}.

For our purposes, the \ac{GRB} trigger provides a known sky position (accurate to within a few degrees) and approximate arrival time (to within a few minutes) of the \ac{GW} signal, as well as motivating some possible signal models.
Furthermore, in each model the \acp{GW} are emitted by a quadrupolar mass distribution rotating around the \ac{GRB} jet axis.
Since the \ac{GRB} is observed at Earth, this implies the observer is near the system axis, which yields circularly polarized \acp{GW} \cite{Kobayashi:2002ez}.

For training and testing the \ac{MVA} classifier we need to choose a set of simulated \ac{GW} waveforms to generate our signal data set.
Since the expected \ac{GW} emission is not known with certainty (particularly for long \acp{GRB}), we must be careful to avoid training the classifier to find only the waveforms that have been used for training.
To do this we use a combination of different waveform classes, which are described below.

\begin{figure}
    \centering
    \subfigure[\,CSG]{\includegraphics[width=0.2\textwidth]{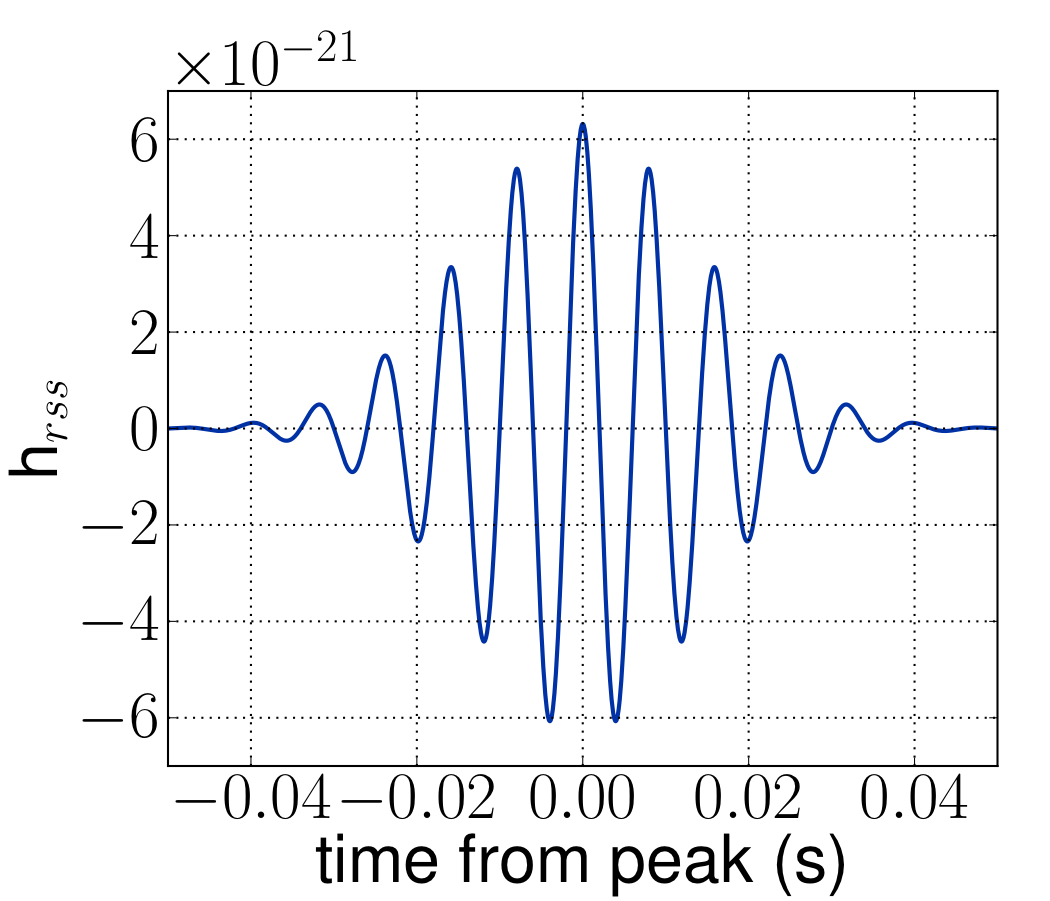}}
    \label{fig:csg_waveform}
    \subfigure[\,inspiral]{\includegraphics[width=0.2\textwidth]{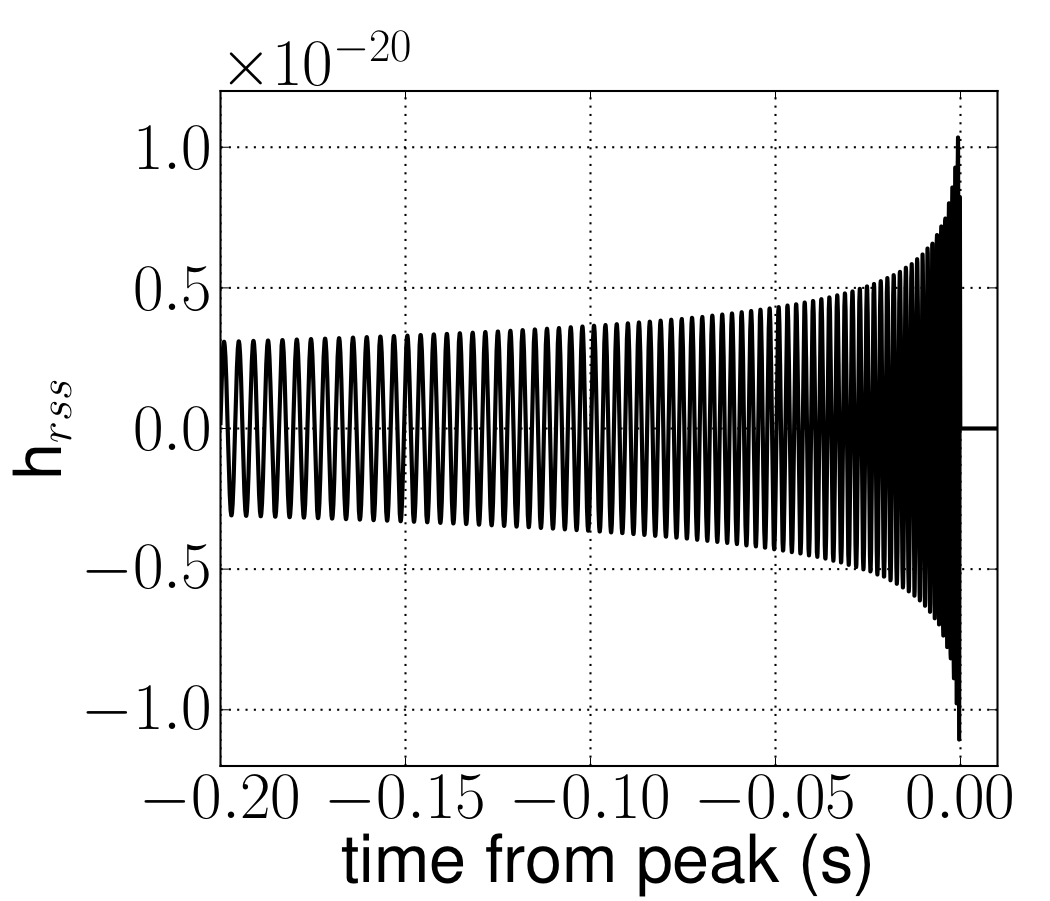}}\\
    \label{fig:insp_waveform}
    \subfigure[\,chirplet]{\includegraphics[width=0.2\textwidth]{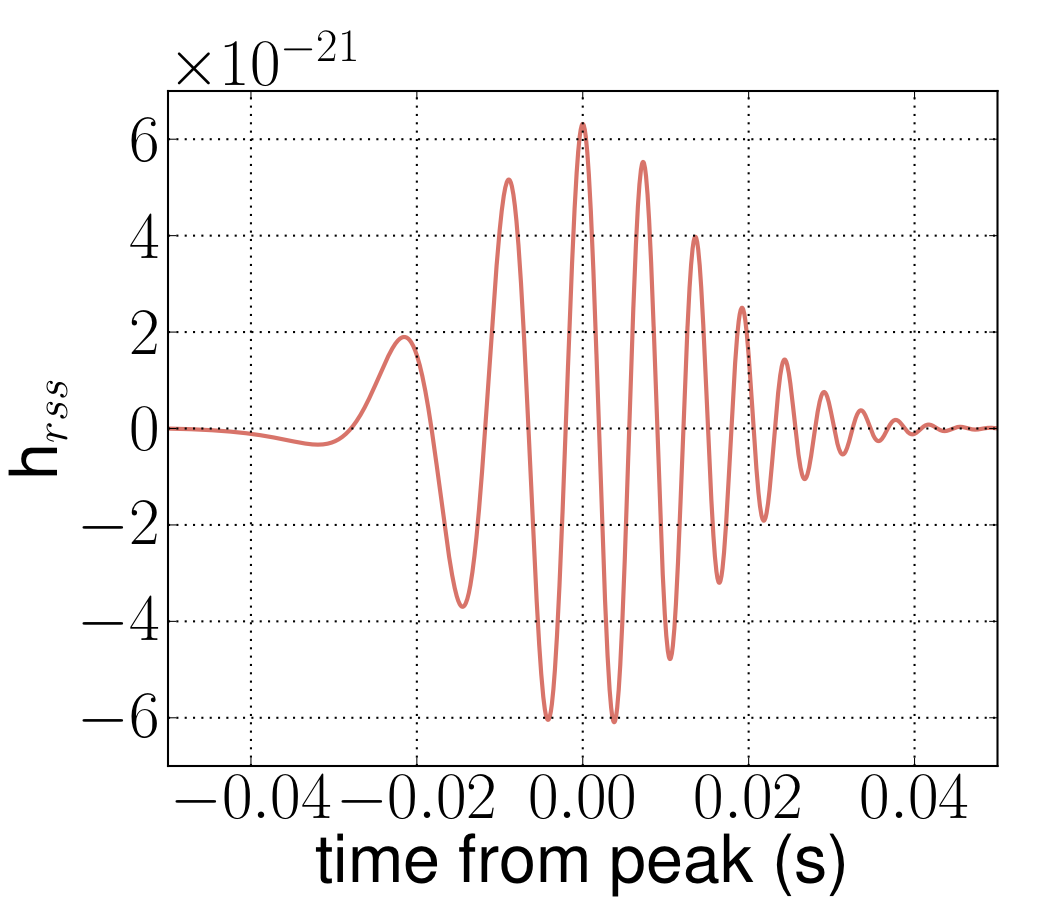}}
    \label{fig:chirplet_waveform}
    \subfigure[\,WNB]{\includegraphics[width=0.2\textwidth]{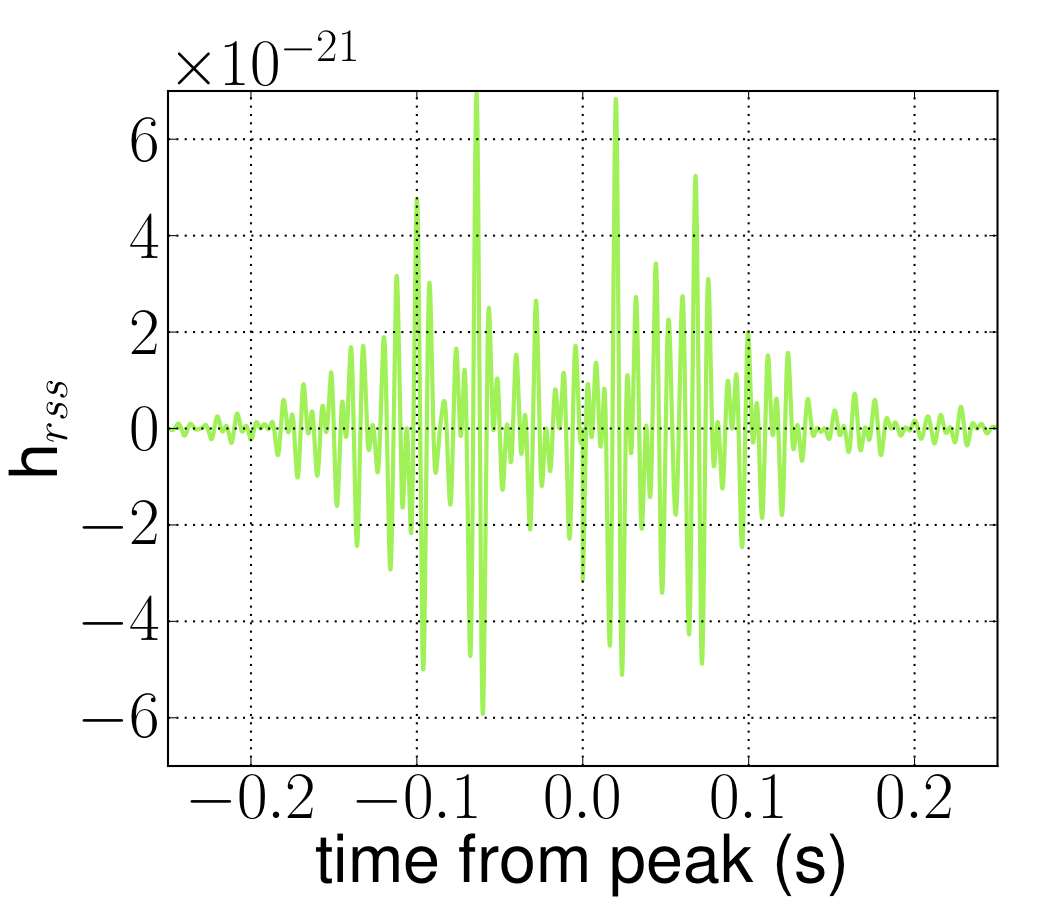}}
    \label{fig:wnb_waveform}
  \caption{\label{fig:waveforms}Time series of waveforms used for signal injections.
  For clarity, only one of the two polarisations is plotted.}
\end{figure}

\begin{description}
\item[circular sine-Gaussians (CSGs):]

\acp{CSG} are circularly polarized, Gaussian-modulated sinusoids with a fixed central frequency and quality factor (number of cycles); see Fig.~\ref{fig:csg_waveform}.
This simple {\em ad hoc} waveform is a standard choice for evaluating the sensitivity of burst searches, and is a special case of the chirplets, which are described below.

\item[binary neutron star inspirals (BNS):]

The binary neutron star progenitor model for short \acp{GRB} implies an associated ``chirp'' signal in \acp{GW} which can be accurately modelled using a Post-Newtonian expansion \cite{Blanchet96}.
See Fig.~\ref{fig:insp_waveform} for an example.
Since the \textsc{X-Pipeline} analysis is not sensitive to the precise morphology, we use the approximation that is quadrupolar in amplitude and 2\,PN in phase and frequency, and cut off the inspiral at the earlier of the coalescence time or the time that the phase second derivative becomes negative.

\item[chirplets:]

Chirplets are a generalisation of the \ac{CSG} waveforms with a non-zero chirp parameter that causes the instantaneous frequency to increase or decrease linearly with time.
See Fig.~\ref{fig:chirplet_waveform} for an example.

\item[white noise bursts (WNBs):]

White noise bursts are stochastic signals -- bursts of Gaussian noise which are white over a frequency band $[f\sub{low}, f\sub{low}+\delta f]$ and which have a Gaussian time profile with decay time $\tau$.
See Fig.~\ref{fig:wnb_waveform} for an example.
\end{description}

The incident sky position is distributed over the \ac{GRB} sky uncertainty region following a Fisher distribution \cite{Was:2012ey}.
The signal arrival time is distributed uniformly over the interval [$t\sub{GRB} - 120\,$s, $t\sub{GRB} + 60\,$s], known as the {\em on-source} window.
Here $t\sub{GRB}$ is the time of the \ac{GRB} trigger; this on-source window is wide enough to encompass most plausible scenarios of \ac{GW} emission associated with \acp{GRB}.
The polarisation angle is uniformly distributed over [0,$\pi$].
For the \acp{CSG} and chirplets, the central frequency is distributed uniformly over the search band, 64\,Hz to 500\,Hz, which is the most sensitive frequency band of the LIGO and Virgo detectors.
The signal decay rate $\tau$ is uniformly distributed between the minimum ($1/4$\,s) and maximum ($1/128$\,s) time resolutions searched by \textsc{X-Pipeline}.
The chirp parameter is distributed uniformly between the values which half or double the central frequency in the time interval from $-\tau$ to $\tau$ about the peak time.
The \ac{BNS} signals use a fixed mass of $1.35\,M_{\odot}$ for each of the components of the binary, and an inclination angle between $0^\circ$ and $30^\circ$.
The \ac{WNB} waveforms are constructed with fixed values $f\sub{low}=50\,\mathrm{Hz}$, $\delta f=100\,\mathrm{Hz}$ and decay time $\tau=0.1\,\mathrm{s}$.

The \ac{BNS} waveforms are physically motivated signal models.
While the other waveforms are {\em ad hoc}, the \acp{CSG} are a standard waveform class for evaluating the sensitivity of \ac{GWB} searches.
Therefore we choose to use a combination of \ac{BNS} and \ac{CSG} waveforms for our default training signal set.
The \ac{WNB} and \ac{chirplet} waveforms are used to test the robustness of the analysis, as described in Section~\ref{sec:robustness_results}.
In particular, the \ac{WNB} model, being stochastic, provides a rigorous test of the ability of \ac{MVA} to detect signals of {\em a priori} unknown shape.

\begin{figure}
    \centering
    \subfigure[\,CSG]{\includegraphics[width=0.2\textwidth]{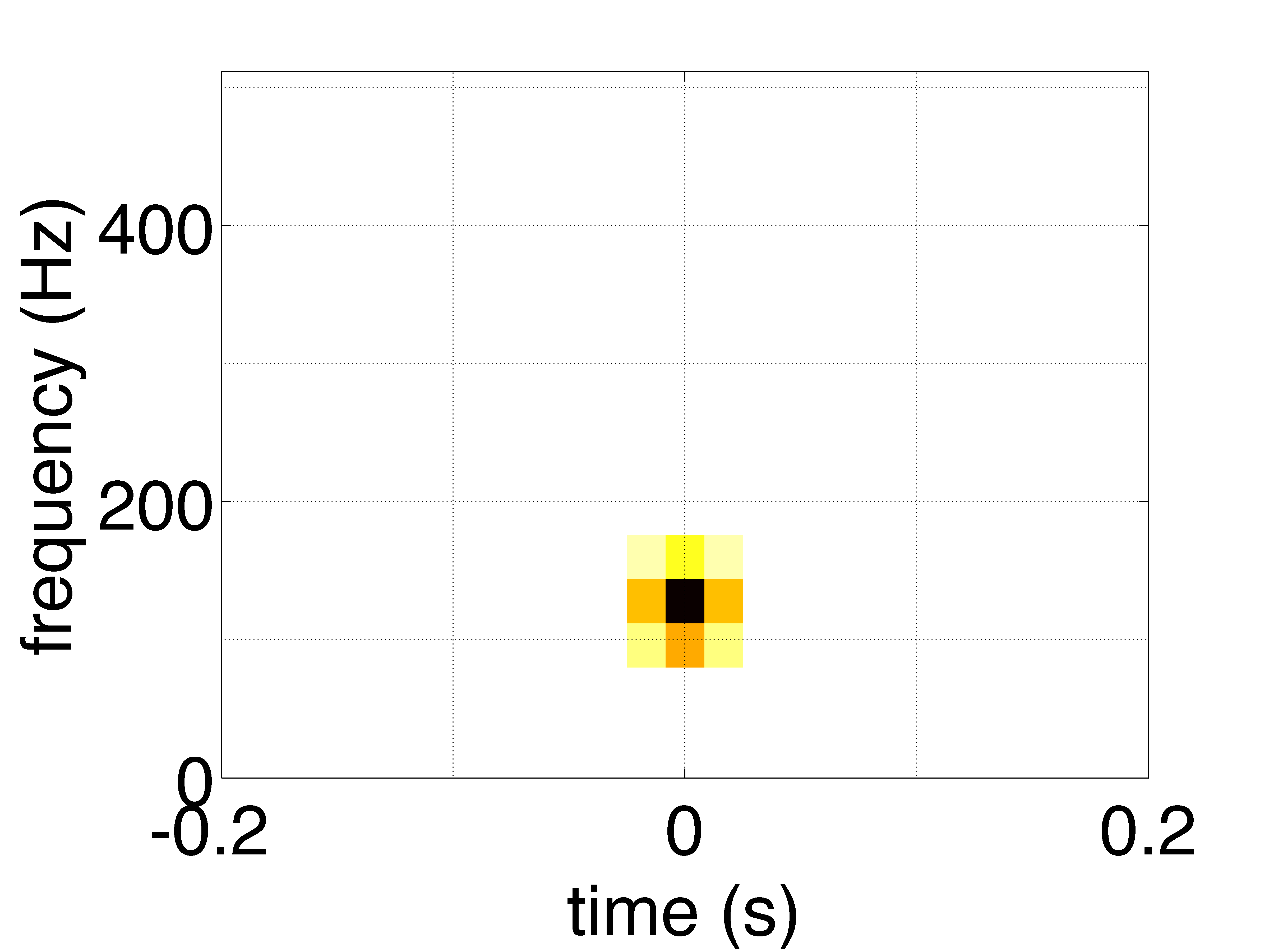}}
    \label{fig:csg_spectrogram}
    \subfigure[\,inspiral]{\includegraphics[width=0.2\textwidth]{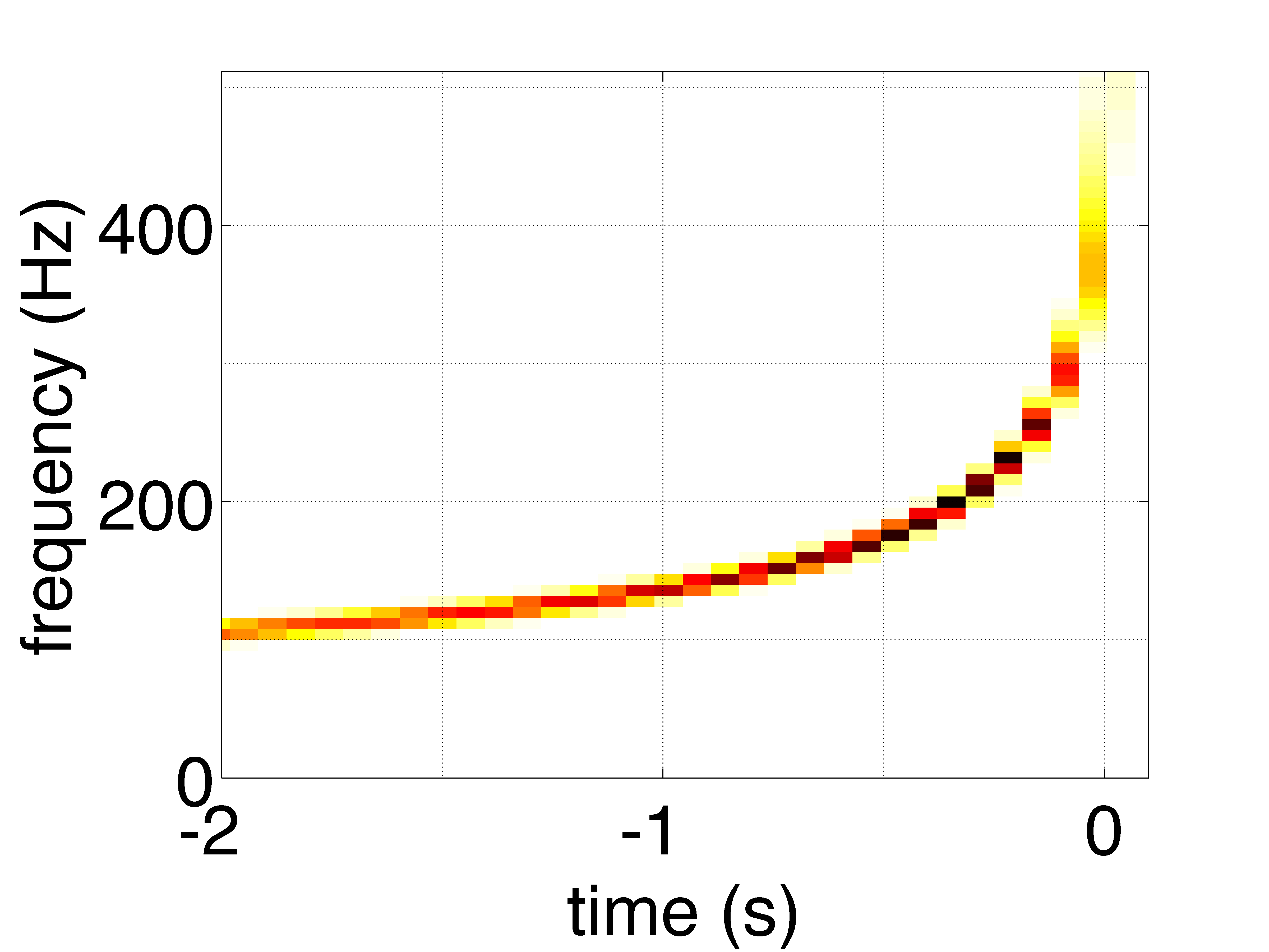}}\\
    \label{fig:insp_spectrogram}
    \subfigure[\,chirplet]{\includegraphics[width=0.2\textwidth]{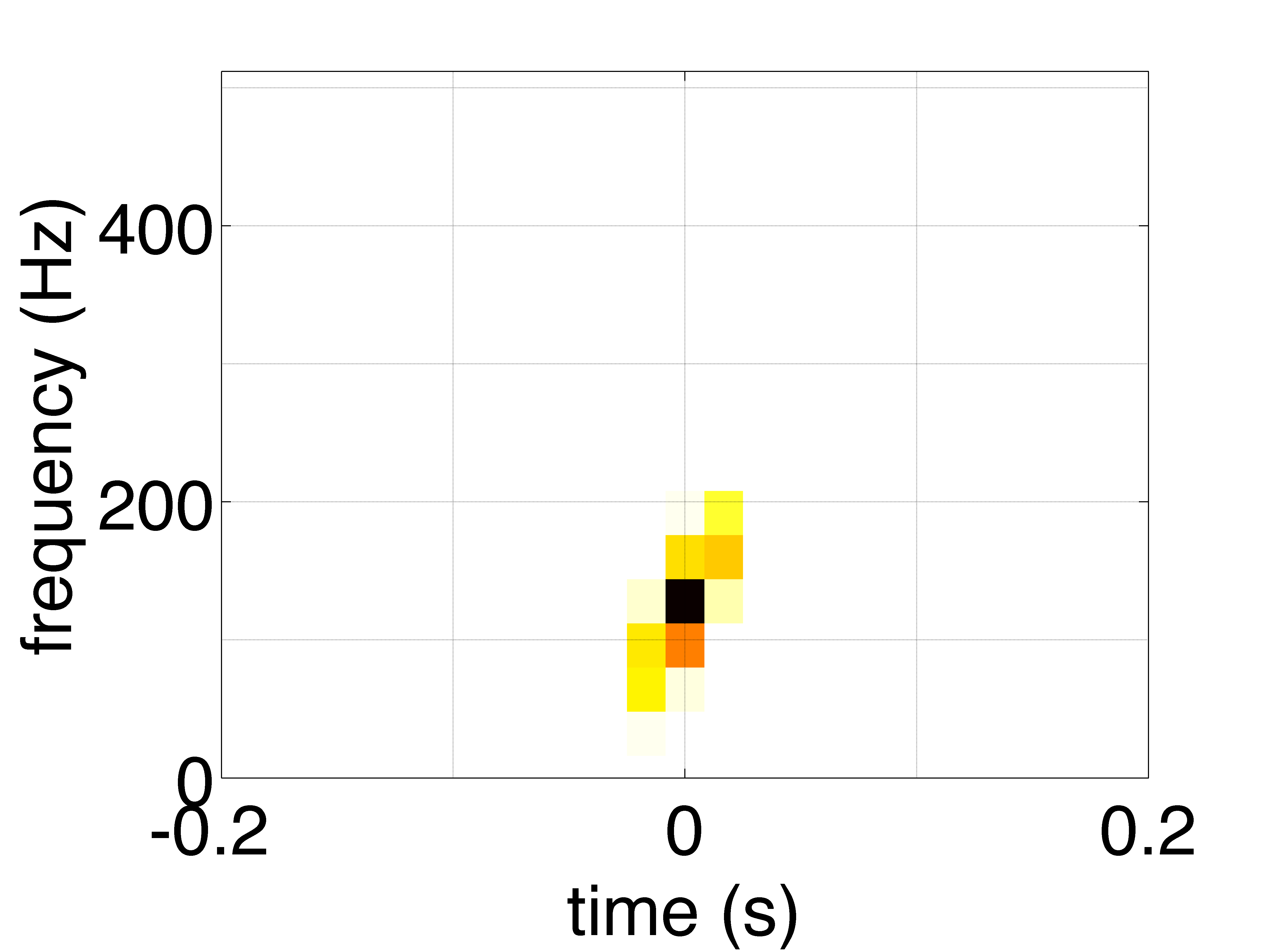}}
    \label{fig:chirplet_spectrogram}
    \subfigure[\,WNB]{\includegraphics[width=0.2\textwidth]{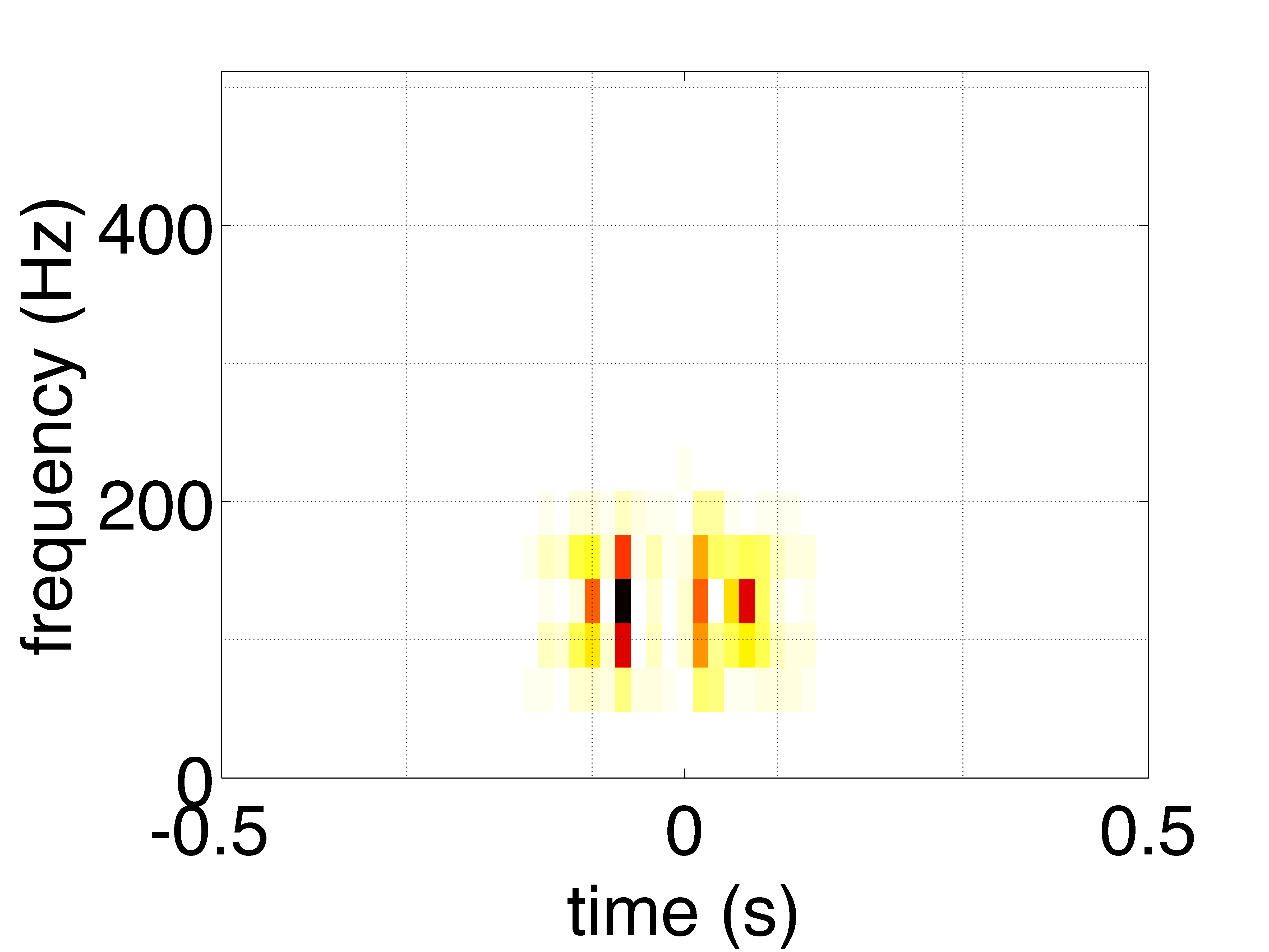}}
    \label{fig:wnb_spectrogram}
  \caption{\label{fig:waveformsTF}Time-frequency spectrograms of waveforms used for signal injections.
  For clarity, only one of the two polarisations is plotted.}
\end{figure}

\section{MVA Performance}\label{sec:results}

We test the efficacy of \ac{MVA} for \ac{GW} burst detection by performing a standard analysis of the type used to search for \acp{GW} from \acp{GRB}.
First, \textsc{X-Pipeline} is used to process the data around the time and sky position of a (simulated) \ac{GRB} trigger.
The sensitivity of the analysis is characterised by the minimum amplitude at which at least 50\% of simulated signals survive the analysis cuts and have $p$ values of 1\% or less, as discussed in Section~\ref{sec:xpipeline}.
This amplitude is denoted by $h\sub{rss}^{50\%}$.
The \ac{BDT} classifier is then applied to the events recorded by \textsc{X-Pipeline} to re-evaluate the significance of each event.
The procedure of cut threshold tuning and sensitivity estimation is then repeated using the \ac{BDT} measure of significance to rank events.
The relative performance is measured as the ratio of $h\sub{rss}^{50\%}$ for the standard \textsc{X-Pipeline} analysis and the \ac{BDT} analysis.
A ratio greater than unity indicates that the \ac{BDT} analysis is more sensitive to a particular waveform type than the standard \textsc{X-Pipeline} analysis.

To verify that the performance improvement of \ac{MVA} is robust, we repeat the \ac{GRB} analysis for a number of different scenarios.
Specifically, we test different \ac{GRB} sky positions covering a range of network sensitivities, and both large and small sky position uncertainty regions.
We also repeat the analysis for a period of particularly poor data quality, and using simulated Gaussian noise to approximate ideal data quality.
We find that the relative improvement of \ac{BDT} to \textsc{X-Pipeline} is consistent across all of these scenarios.
We also explore the effect of training using only two types of waveform (\acp{CSG}, \acp{BNS}) or all four types of waveform.
We find that even when searching for signal types that are not included in the training set, the \ac{BDT} analysis is consistently at least as sensitive as the \textsc{X-Pipeline} analysis, and typically more sensitive.
Furthermore, in all cases we find that after processing events with \ac{BDT}, the \textsc{X-Pipeline} background rejection tests do not improve the sensitivity further; i.e., the \ac{BDT} has effectively incorporated the signal/background discrimination power of the \textsc{X-Pipeline} background rejection test.
Since the test typically requires some assumption about the signal polarisation (in this study we assume circular polarisation), the replacement of the test by \ac{BDT} actually broadens the range of signals to which the analysis is sensitive.

The following subsections describe each of these tests in turn.
We give a full table of the $h\sub{rss}^{50\%}$ results for all analyses and all waveforms in Table~\ref{tab:results}.

\begin{table*}
  \begin{center}
    \begin{tabular}{r|c|c|c|c}
    \hline
    \hline
        Test name & UTC time & Right ascension & Declination & Sky position uncertainty \\
    \hline
    \hline
        GRB\,060223A (default) & 2006-02-23 06:04:23 & $55.19^{\circ}$ & $-17.13^{\circ}$ & $0.03^{\circ}$ \\
    \hline
        sky position 1 & 2006-02-23 06:04:23 & $299.56^{\circ}$ &  $44.16^{\circ}$ & $0.03^{\circ}$ \\
        sky position 2 & 2006-02-23 06:04:23 & $311.02^{\circ}$ &  $32.70^{\circ}$ & $0.03^{\circ}$ \\
        sky position 3 & 2006-02-23 06:04:23 & $345.40^{\circ}$ &  $-1.67^{\circ}$ & $0.03^{\circ}$ \\
        sky position 4 & 2006-02-23 06:04:23 &  $31.23^{\circ}$ & $-47.51^{\circ}$ & $0.03^{\circ}$ \\
    \hline
        large sky position uncertainty 1 & 2006-02-23 06:04:23 & $345.40^{\circ}$ &  $-1.67^{\circ}$ & $9.0^{\circ}$ \\
        large sky position uncertainty 2 & 2006-02-23 06:04:23 &  $31.23^{\circ}$ & $-47.51^{\circ}$ & $9.0^{\circ}$ \\
    \hline
        highly non-Gaussian background & 2007-06-20 03:05:40 & $319.52^{\circ}$ & $-57.67^{\circ}$ & $0.03^{\circ}$ \\
    \hline
        detection challenge & 2007-09-22 03:05:40 & $33.44^{\circ}$ & $16.94^{\circ}$ & $0.03^{\circ}$ \\
    \hline
    \hline
    \end{tabular}
  \caption{\label{tab:test_parameters} Trigger parameters used during test analyses.
  The waveform robustness and Gaussian noise tests used the default GRB\,060223A parameters.}
  \end{center}
\end{table*}

\subsection{GRB\,060223A analysis}
\label{sec:default_results}

For our baseline test we perform an analysis using the parameters (time, sky position) of GRB\,060223A, as given in Table~\ref{tab:test_parameters}.
GRB\,060223A was detected by the {\em Swift} satellite \citep{swift04} during a period of operation of the LIGO H1, LIGO L1, and Virgo V1 detectors, and localised by {\em Swift} to a well-defined sky position.
We generate signal events by adding simulated \ac{CSG} and \ac{BNS} signals to the three minute on-source window $t\sub{GRB} - 120\,$s, $t\sub{GRB}+60\,$s around the \ac{GRB}.
We generate background events by analysing a three-hour off-source window surrounding the \ac{GRB} time.
These events were split randomly into two sets for training and testing the \ac{BDT}.

As can be seen in Fig.~\ref{fig:results}, for \ac{CSG} signals the \ac{BDT} analysis gives a substantial improvement in sensitivity -- of order $50\%$ -- over the standard \textsc{X-Pipeline} analysis.
However, there is no significant improvement in the sensitivity to \ac{BNS} signals (differences of order 5\% are not statistically significant).

\subsection{Sky position}
\label{sec:sky_position_results}

To verify that the results of the \ac{BDT}--\textsc{X-Pipeline} comparison are robust, we repeat the test for a variety of other cases.
First, we vary the sky position of the \ac{GRB} trigger.
We test four additional sky positions, as listed in Table~\ref{tab:test_parameters}.
These positions were chosen to cover a range of different relative detector network sensitivities \cite{Was:2012ey}.

As can be seen in Fig.~\ref{fig:results}, for \ac{CSG} waveforms the \ac{BDT} analysis gives a consistent improvement in sensitivity of $30-50\%$, for all tested sky positions, compared to the standard \textsc{X-Pipeline} analysis.
Again there is no significant change in the sensitivity to \ac{BNS} signals.

\subsection{Large sky position uncertainty}
\label{sec:sky_pos}

The previous tests have assumed the \ac{GRB} sky position to be known to high accuracy ($\ll 1^\circ$).
By contrast, \acp{GRB} detected by the GBM instrument on the Fermi satellite have relatively large sky location systematic uncertainties of a few degrees~\cite{Briggs09} and statistical errors of up to $\sim$10 degrees.
This requires analysing the \ac{GW} data over a grid of trial sky positions covering the error region~\cite{Was:2012ey}.
We test the performance of the \ac{BDT} analysis in this scenario using two different sky positions with sky position uncertainties of $\approx 9^{\circ}$ (see Table~\ref{tab:test_parameters}), which is typical for Fermi-GBM \acp{GRB} \cite{Meegan:2009qu}.

As can be seen in Fig.~\ref{fig:results}, the \ac{BDT} performance is consistent with previous tests: for \ac{CSG} waveforms \ac{BDT} improves the sensitivity by $40-50\%$, with no significant change in the sensitivity to \ac{BNS} signals.

\subsection{Highly non-Gaussian (glitchy) background}
\label{sec:glitchy_data}

Excess power noise transients can be introduced into the detector data streams by a wide range of known and unknown sources.
These glitches are artefacts of the detectors and can be difficult to distinguish from real weak signals.
To test the performance of the \ac{BDT} analysis, we analyse a trigger which is at a time of unusually poor data quality.

As can be seen in Fig.~\ref{fig:results}, for \ac{CSG} waveforms the \ac{BDT} analysis again gives a $\sim$30\% improvement in sensitivity, with no notable change for \ac{BNS} signals compared to the standard \textsc{X-Pipeline} analysis.

\subsection{Gaussian background}
\label{sec:gaussianity_results}

As a best-case scenario, the performance of the \ac{BDT} analysis was tested using simulated Gaussian noise with a spectral density coloured to match that of the real detector noise at the time of our default GRB\,060223A trigger.
All other parameters are kept the same as in the default analysis.

As can be seen in Fig.~\ref{fig:results}, for \ac{CSG} waveforms the \ac{BDT} analysis gives an improvement in sensitivity of $40\%$ compared to the standard \textsc{X-Pipeline} analysis.
There is no notable change in the sensitivity to \ac{BNS} signals.

\subsection{Waveform robustness}
\label{sec:robustness_results}

In \ac{GW} burst searches the signal waveform is usually not known \textit{a priori}.
It is therefore of the utmost importance to verify that \ac{MVA} is able to detect waveforms with morphologies that differ from those used for training; at the very least, \ac{MVA} should not have \textit{worse} sensitivity for unknown waveforms than the standard analysis.
We study this issue by repeating our analysis using different waveform sets for training and testing.
Specifically, we evaluate the \ac{BDT} performance for detecting chirplet and \ac{WNB} waveforms in two cases: one in which the \ac{BDT} is trained using \ac{CSG} and \ac{BNS} signals \textit{only} (and not chirplet or \ac{BNS} signals) and again after training on all four waveform types (\ac{CSG}, \ac{BNS}, chirplet, \ac{WNB}).
We refer to these as the two-waveform and four-waveform robustness tests.

\begin{figure}
  \begin{center}
    \includegraphics[width=0.5\textwidth]{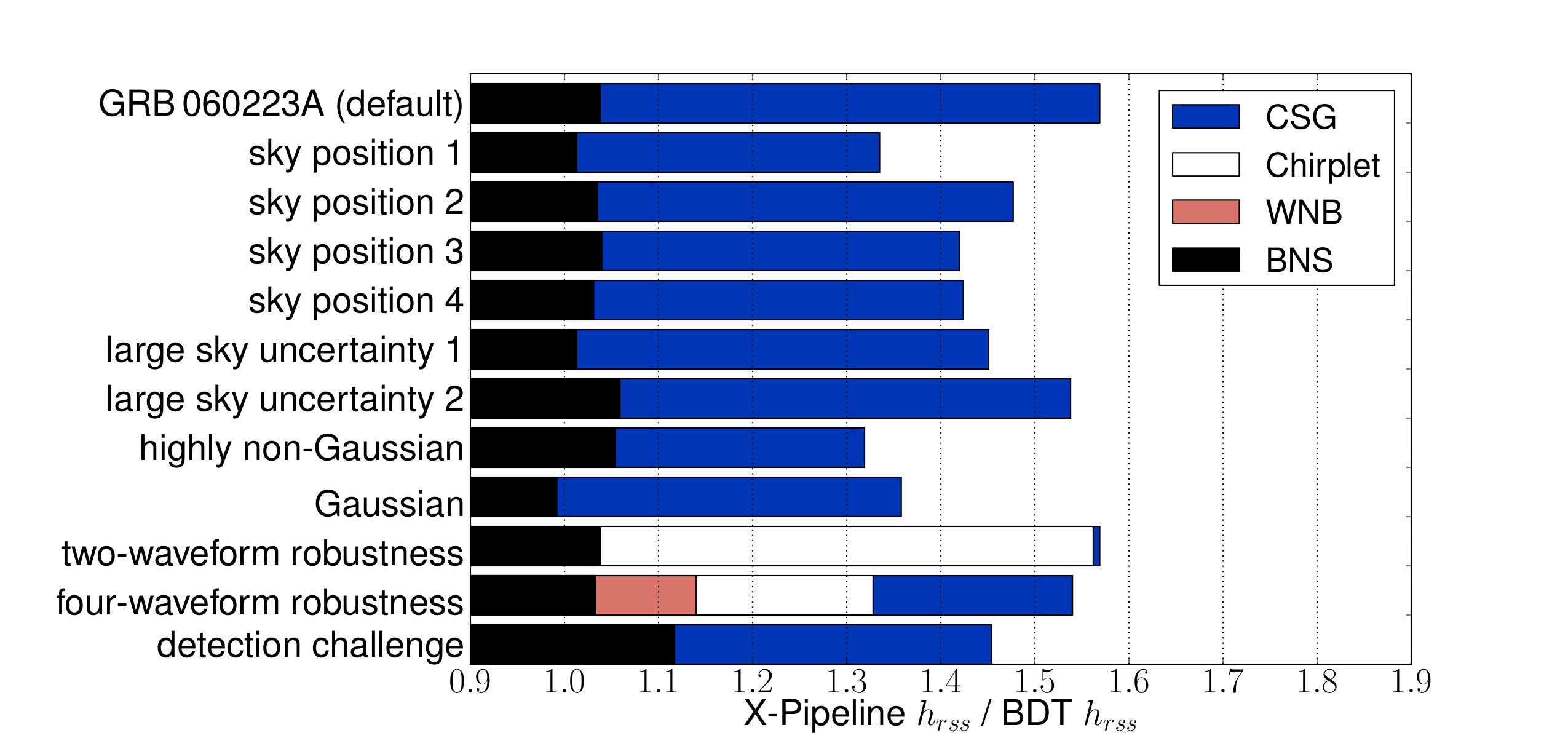}
    \caption{\label{fig:results}Ratio of the minimum-detectable signal amplitudes $h_\mathrm{rss}^{50\%}$ for the standard \textsc{X-Pipeline} analysis and the \ac{BDT} analysis for each of the signal waveforms and scenarios described in Section~\ref{sec:results}.
    Ratios greater than 1 indicate the \ac{BDT} analysis is more sensitive than the standard \textsc{X-Pipeline} analysis.
(Equivalently, for a fixed signal
amplitude the distance reach of the \ac{BDT} search is greater than that of the
\textsc{X-Pipeline} search by this ratio.)
    \ac{CSG} performance is given in blue (dark grey), \ac{BNS} performance is given in black, \ac{chirplet} performance is given in white, and \ac{WNB} performance is given in red (light grey).
    The standard \textsc{X-Pipeline} analysis is unable to recover \ac{WNB} signals in the robustness two-waveform test, so the sensitivity ratio is ill-defined in this case. All $h_\mathrm{rss}^{50\%}$ values of results can be found in Table~\ref{tab:results}.}
  \end{center}
\end{figure}

Fig.~\ref{fig:results} shows that in the two-waveform test (training on \ac{CSG} and \ac{BNS} only) the \ac{BDT} analysis shows the same performance for \ac{CSG} and \ac{BNS} as was seen in the default GRB\,060223A analysis.
This is expected, as the tests are identical as far as these waveforms are concerned.
However, \ac{BDT} also gives an improvement in sensitivity of order $50\%$ for \ac{chirplet} waveforms compared to the standard \textsc{X-Pipeline} analysis.
This implies that the \acp{CSG} and chirplets are sufficiently similar in terms of a time-frequency analysis that an \ac{MVA} trained to detect one can detect the other.
More surprising is the \ac{BDT} performance for \acp{WNB}.
These waveforms are not detectable by the standard \textsc{X-Pipeline} analysis.
This happens because the two \ac{GW} polarisations are uncorrelated for a \ac{WNB}, whereas the \textsc{X-Pipeline} background rejection test applied to the signal stream (discussed in Section~\ref{sec:xpipeline}) assumes the two polarisations are related by $90^\circ$ phase shift, as expected for a circularly polarised signal.
The \ac{BDT} analysis is able to recover these waveforms, albeit with an $h_\mathrm{rss}^{50\%}$ value about twice as high as for the case of training with \acp{WNB} (discussed below).

The four waveform robustness test used independent samples of all four waveform types (\ac{CSG}, \ac{BNS}, \ac{chirplet}, and \ac{WNB}) for both training and testing.
From Fig.~\ref{fig:results} we again see the same performance from \ac{BDT} for the \ac{CSG} and \ac{BNS} signals.
However, with training extended to include chirplet and \ac{WNB} waveforms, we see slightly \textit{less} improvement of sensitivity to chirplets (only $30\%$ compared to the standard \textsc{X-Pipeline} analysis).
This is partly due to a small improvement in the sensitivity of the \textsc{X-Pipeline} analysis to chirplets when they are included in the training.
However, most of the change is due to a decrease in sensitivity of the \ac{BDT} analysis ($\sim$\,15\% drop) from the two-waveform case; we attribute this to the inclusion of \acp{WNB} in the training.
The classifier in this case finds a compromise between the sensitivity to circularly polarised signals and unpolarised signals in the training.
This can be seen as the sensitivity to \acp{WNB} is dramatically improved, by more than a factor of two compared to the two-waveform \ac{BDT} analysis.
The standard \textsc{X-Pipeline} analysis can also detect \acp{WNB} when trained with all four waveform types; in this case the automated background rejection tuning places less emphasis on the tests that assume circular polarisation and more on the polarisation-independent null stream test.
We find a net sensitivity improvement of $\sim\,15\%$ by \ac{BDT} relative to \textsc{X-Pipeline} for \acp{WNB} when training includes these waveforms.

\subsection{Detection challenge case}
\label{sec:detection_results}

\begin{table*}
\begin{center}
\begin{tabular}{l|c|c|c|c}
\hline
\hline
    Analysis & waveform & \textsc{X-Pipeline} $h\sub{rss}^{50\%}$ & \ac{BDT} $h\sub{rss}^{50\%}$ & ratio \\
\hline
\hline
    GRB\,060223A   & \ac{CSG}      & $4.90\times10^{-22}$ & $3.12\times10^{-22}$ & $1.569$ \\
                   & \ac{BNS}      & $1.09\times10^{-21}$ & $1.05\times10^{-21}$ & $1.038$ \\
\hline
    sky position 1 & \ac{CSG}      & $8.70\times10^{-22}$ & $6.51\times10^{-22}$ & $1.335$ \\
                   & \ac{BNS}      & $2.05\times10^{-21}$ & $2.02\times10^{-21}$ & $1.013$ \\
\hline
    sky position 2 & \ac{CSG}      & $6.43\times10^{-22}$ & $4.36\times10^{-22}$ & $1.477$ \\
                   & \ac{BNS}      & $1.42\times10^{-21}$ & $1.37\times10^{-21}$ & $1.035$ \\
\hline
    sky position 3 & \ac{CSG}      & $4.07\times10^{-22}$ & $2.86\times10^{-22}$ & $1.420$ \\
                   & \ac{BNS}      & $9.08\times10^{-22}$ & $8.73\times10^{-22}$ & $1.040$ \\
\hline
    sky position 4 & \ac{CSG}      & $3.91\times10^{-22}$ & $2.74\times10^{-22}$ & $1.424$ \\
                   & \ac{BNS}      & $9.03\times10^{-22}$ & $8.76\times10^{-22}$ & $1.031$ \\
\hline
    large sky position uncertainty 1
                   & \ac{CSG}      & $4.19\times10^{-22}$ & $2.89\times10^{-22}$ & $1.451$ \\
                   & \ac{BNS}      & $9.39\times10^{-22}$ & $9.26\times10^{-22}$ & $1.013$ \\
\hline
    large sky position uncertainty 2
                   & \ac{CSG}      & $4.04\times10^{-22}$ & $2.63\times10^{-22}$ & $1.538$ \\
                   & \ac{BNS}      & $9.21\times10^{-22}$ & $8.70\times10^{-22}$ & $1.059$ \\
\hline
    highly non-Gaussian background
                   & \ac{CSG}      & $5.70\times10^{-22}$ & $4.32\times10^{-22}$ & $1.319$ \\
                   & \ac{BNS}      & $1.38\times10^{-21}$ & $1.31\times10^{-21}$ & $1.054$ \\
\hline
    Gaussian background
                   & \ac{CSG}      & $4.48\times10^{-22}$ & $3.30\times10^{-22}$ & $1.358$ \\
                   & \ac{BNS}      & $1.01\times10^{-21}$ & $1.02\times10^{-21}$ & $0.992$ \\
\hline
    two-waveform robustness
                   & \ac{CSG}      & $4.90\times10^{-22}$ & $3.12\times10^{-22}$ & $1.569$ \\
                   & \ac{BNS}      & $1.09\times10^{-21}$ & $1.05\times10^{-21}$ & $1.038$ \\
                   & chirplet      & $5.03\times10^{-21}$ & $3.22\times10^{-22}$ & $1.562$ \\
                   & \ac{WNB}      & nan                  & $1.65\times10^{-21}$ & nan \\
\hline
    four-waveform robustness
                   & \ac{CSG}      & $4.91\times10^{-22}$ & $3.19\times10^{-22}$ & $1.540$ \\
                   & \ac{BNS}      & $1.10\times10^{-21}$ & $1.06\times10^{-21}$ & $1.033$ \\
                   & chirplet      & $4.81\times10^{-22}$ & $3.62\times10^{-22}$ & $1.328$ \\
                   & \ac{WNB}      & $7.45\times10^{-22}$ & $6.54\times10^{-22}$ & $1.140$ \\
\hline
    detection challenge case
                   & \ac{CSG}      & $4.92\times10^{-22}$ & $3.38\times10^{-22}$ & $1.454$ \\
                   & \ac{BNS}      & $7.65\times10^{-22}$ & $6.85\times10^{-22}$ & $1.117$\\
\hline
\hline
\end{tabular}
\caption{\label{tab:results}
Sensitivity of the standard \textsc{X-Pipeline} and \ac{BDT}-augmented analyses
for each test scenario and waveform type.  $h\sub{rss}^{50\%}$ (Hz$^{-1/2}$) is
the minimum amplitude for which at least 50\% of simulated signals survive the
analysis cuts and have FAP of 1\% or less.  The last column is the ratio of
$h\sub{rss}^{50\%}$ for \textsc{X-Pipeline} and \ac{BDT}; values greater than
unity indicate that \ac{BDT} is more sensitive.  Equivalently, for a fixed signal
amplitude the distance reach of the \ac{BDT} search is greater than that of the
\textsc{X-Pipeline} search by this ratio.
}
\end{center}
\end{table*}

Recent science runs of the LIGO and Virgo detectors have included a ``blind injection challenge'' wherein a small number of simulated signals are secretly added to the data via the interferometer control systems \cite{lowMassS5y2,2010ApJ...715.1453A,Abadie:2012cf,lowMassS6}.
These signals are used to test the analysis procedures.
Our final \ac{MVA} test is to analyse one of these signals, to demonstrate that the improvement in sensitivity extends to false-alarm rates low enough to permit a detection claim at the $3\sigma$ level.

\begin{figure}
  \includegraphics[width=0.4\textwidth]{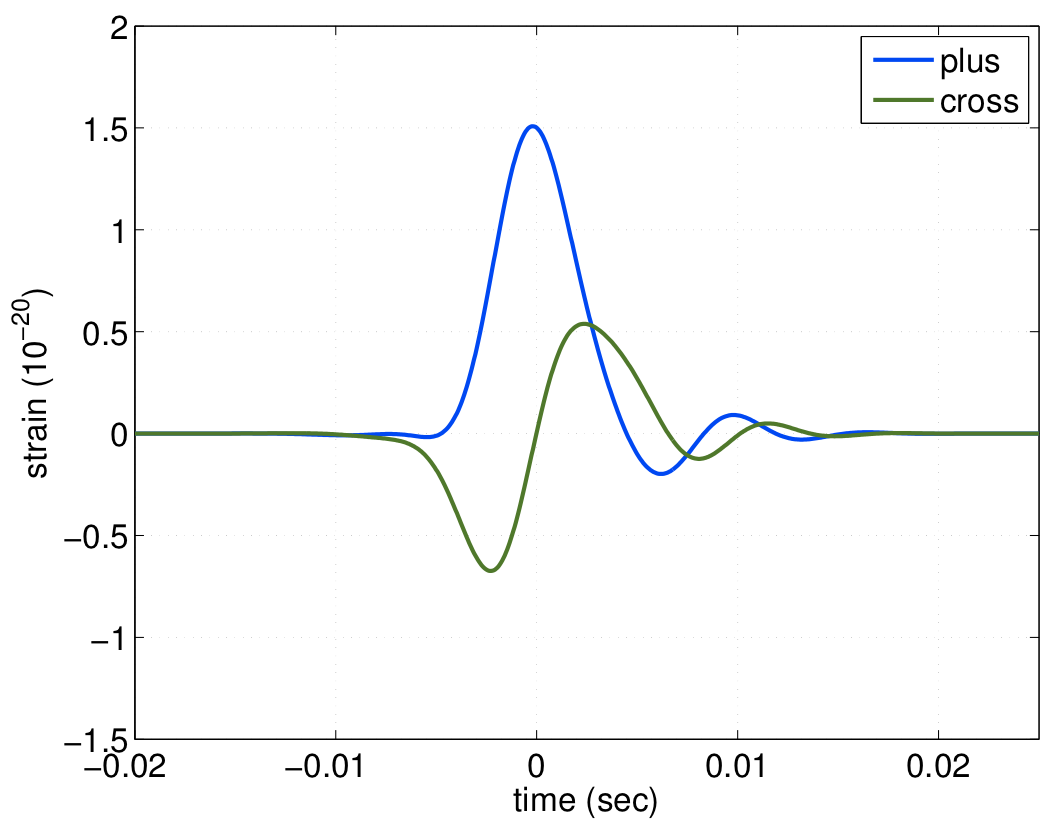}
  \caption{\label{fig:equinox}
  Time series of the ``equinox event'' signal in the detection challenge test.
}
\end{figure}

For this test we select the ``equinox event'', an injection performed on 22 September 2007.  The simulated waveform was approximately a single-cycle sine-Gaussian with a central frequency of approximately 60\,Hz and an amplitude of $h_\mathrm{rss}=1.0\times10^{-21}\mathrm{Hz}^{-1/2}$; see Fig.~\ref{fig:equinox}.
The relative amplitudes of the plus and cross polarisations were consistent with an inclination angle of approximately $30^\circ$.
The sky position is shown in Table~\ref{tab:test_parameters}.

We analysed this injection using the standard \ac{GRB} procedure; i.e., assuming the sky position and approximate time of the event were known \textit{a priori} due to observation of an electromagnetic counterpart.
In our previous tests we evaluated the minimum detectable signal amplitude at a fixed false alarm probability of 1\%.
This follows the standard use of \textsc{X-Pipeline} in \ac{GRB} searches \cite{burstGrbS5,Abadie:2012cf}.
However, in order to claim the detection of a gravitational-wave signal, much lower false-alarm probabilities are required.
In particular, a $3\sigma$ significance requires a false-alarm probability of $p\le0.0027$.
Furthermore, a typical search includes 100--150 \ac{GRB} triggers, which must be accounted for in the trials factor.
A $3\sigma$ significance with 150 trials requires $p\lesssim2\times10^{-5}$ for an individual event.
For this analysis we therefore generate extra background samples and tune the background rejection tests to yield the lowest minimum injection amplitude at a \ac{FAP} of $p=10^{-5}$.
Since the blind injection was not added to the Virgo detector data, we analyse the event using the LIGO H1 and L1 detectors only.
All other analysis parameters are the same as for the GRB\,060223A test, including training and testing with \ac{CSG} and \ac{BNS} waveforms.

\begin{figure}
  \includegraphics[width=0.5\textwidth]{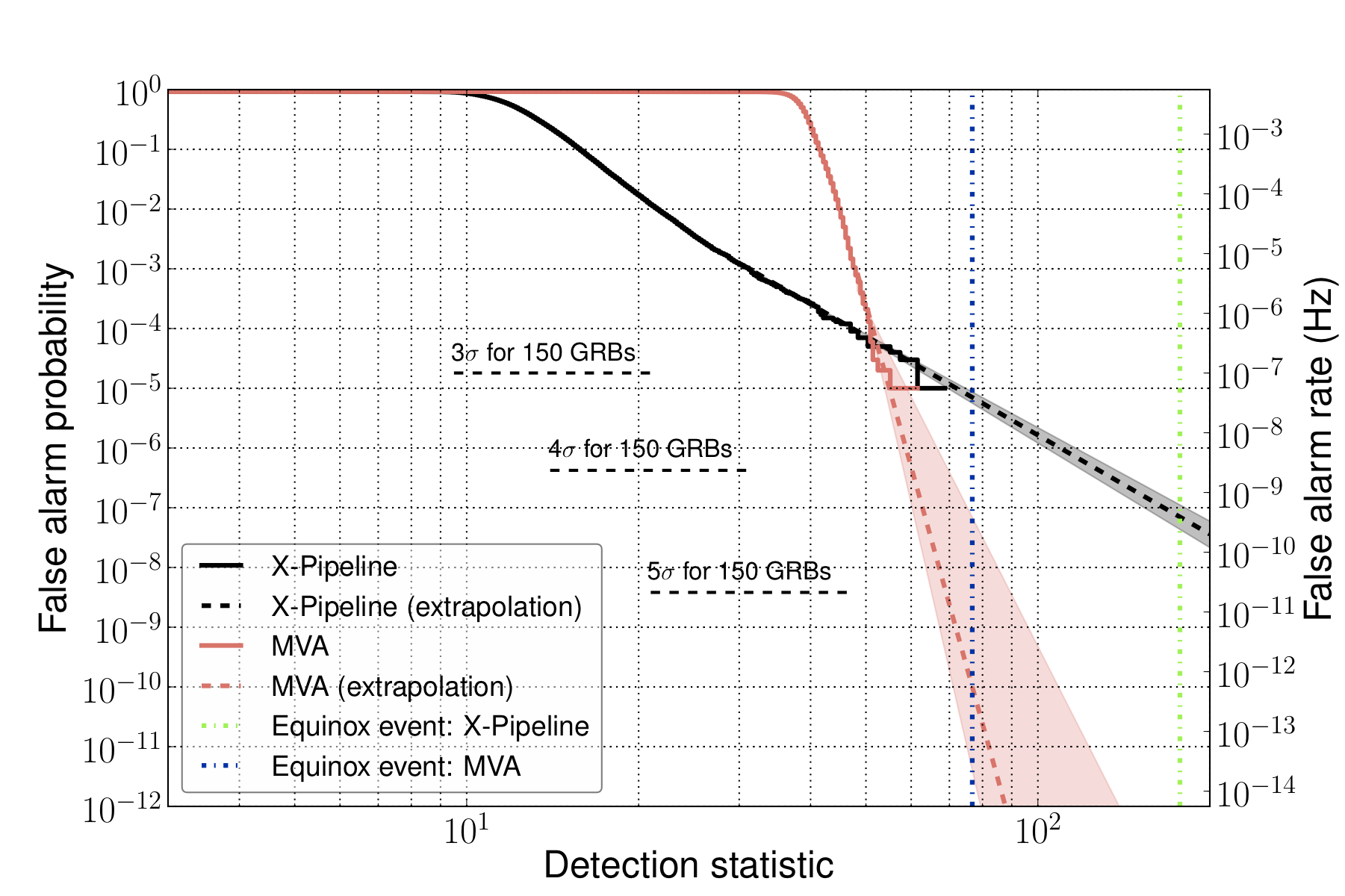}
  \caption{\label{fig:background_hist}Background distributions
  for the \textsc{X-Pipeline} and \ac{BDT} analyses of the equinox event.
  The background distributions are shown as solid lines.  The power-law
  extrapolations are shown as dashed lines, with the shading indicating
  the estimated uncertainty in the extrapolations.
  The value of the detection statistic of the equinox event is shown by the green and blue vertical
  lines for the \textsc{X-Pipeline} and \ac{BDT} analyses respectively.}
\end{figure}

Fig.~\ref{fig:background_hist} shows the cumulative distribution of the detection statistic for the loudest background event per three minute interval (the on-source interval) returned by the standard \textsc{X-Pipeline} analysis and the \ac{BDT} analysis.
Both distributions are consistent with a power-law relationship between false alarm probability and detection statistic down to the lowest false alarm probabilities measured, $p\simeq10^{-5}$.
From Fig.~\ref{fig:results} we can see that the \ac{BDT} analysis gives an improvement in sensitivity compared to the standard \textsc{X-Pipeline} analysis that is consistent with previous tests.
This demonstrates that the benefits of the \ac{BDT} analysis extend down to false alarm rates sufficient for $3\sigma$ detections in \ac{GRB} triggered searches.

Of particular interest is the significance assigned to the equinox event itself.
The vertical lines in Fig.~\ref{fig:background_hist} indicate the value of the detection statistic returned by \textsc{X-Pipeline} and \ac{BDT}.
In both analyses the equinox event is clearly detected, with a significance higher than any of the background events.
In order to estimate the approximate false alarm probability for the equinox event, we extrapolate the background distributions using a best-fit power law.
This yields $p\simeq7\times10^{-8}$ ($5.4\sigma$) for the standard \textsc{X-Pipeline} analysis and $p\simeq1\times10^{-10}$ ($6.5\sigma$) for the \ac{BDT} analysis.
The shaded bands indicate the plausible extrapolations from using a varied number of data points to determine the best-fit parameters, which can be taken as an estimate of the uncertainty in the extrapolation.
While these are estimates, the \ac{BDT} analysis assigns a false alarm probability which is at worst consistent with the \textsc{X-Pipeline} result, and the range of possible extrapolations suggest that the false alarm probability could be significantly lower.

\section{Discussion and Conclusions}\label{sec:summary}

The tests shown in Section~\ref{sec:results} and Fig.~\ref{fig:results} demonstrate that the \ac{BDT}-augmented analysis yields a consistent improvement in sensitivity to some signal types at fixed false-alarm rate with respect to the standard \textsc{X-Pipeline} analysis.
The improvement holds regardless of the incident direction of the signal or its sky location uncertainty, the data quality, and the network of detectors.
Most importantly, the \ac{BDT} analysis is always at least as sensitive as \textsc{X-Pipeline}, even to signals of different morphology to those used in training, and the sensitivity improvement extends down to false alarm probabilities required for detection.

The degree to which \ac{BDT} outperforms the standard \textsc{X-Pipeline} analysis depends on the signal waveform.
For \acp{CSG}, which have compact time-frequency distributions, we find a consistent improvement in sensitivity of $35-55\%$.
By contrast, for \ac{BNS} signals, which are long-duration and have extended time-frequency distributions, the average improvement in sensitivity is only 4\%.
The \ac{BDT} analysis also yields improved sensitivity to \ac{chirplet} and \ac{WNB} waveforms, regardless of whether they were included in the training set or not.
The large increase in sensitivity to \ac{chirplet} waveforms seen in the two-waveform robustness test is likely due to these waveforms being very similar to the \acp{CSG}.
The smaller improvement seen in the four-waveform test is likely due to the classifier compromising performance between the mix of waveforms; the effect of this is a decrease in sensitivity gain for both the \ac{CSG} and \ac{chirplet} waveforms, but a dramatic improvement in sensitivity to \ac{WNB} waveforms.

The robustness of the \ac{BDT} analysis to signals of different morphology from those used in training is crucial, because accurate signal waveforms are not known in most burst searches. (In fact, the \ac{WNB} results from the two-waveform robustness test show that \ac{BDT} can actually improve the sensitivity to \textit{a priori} unknown waveforms by removing the need for \textsc{X-Pipeline}'s polarisation-specific background rejection tests.)
The robustness of \ac{BDT} may be due to the fact that the \ac{MVA} does not have access to the raw \ac{GW} data, but rather only to characteristics passed on by \textsc{X-Pipeline}.
In particular, the only time-frequency information that is available to \ac{BDT} are the time and frequency extent of the event, its peak time and frequency, and the number of time-frequency pixels; no shape information is recorded.
In principle shape information could be used to improve signal/background discrimination, e.g. by recognising the characteristic chirp shape of inspiral signals (see Fig.~\ref{fig:insp_waveform}).
However, this would presumably also make the \ac{MVA} analysis more waveform-specific, and less sensitive to signals not included in the training.  Further study of the waveform dependence of \ac{MVA} analyses is warranted.

\begin{figure}
    \includegraphics[width=0.4\textwidth]{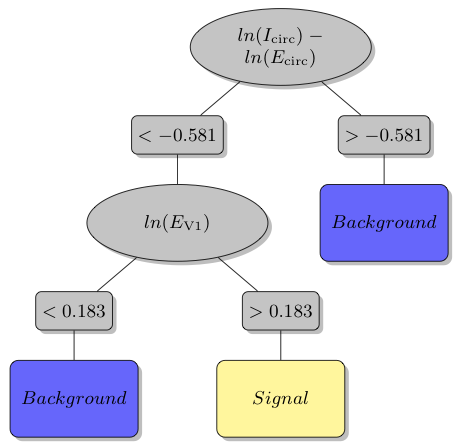}
    \caption{\label{fig:decisiontreeexample1}Example decision tree used in the GRB\,060223A \ac{BDT} analysis.
    Decision nodes are grey ellipses, with thresholds for a branch given in grey rectangles.
    Leaf nodes for signal events are light yellow rectangles and for background events are dark blue rectangles.}
\end{figure}

\begin{figure}
    \includegraphics[width=0.45\textwidth]{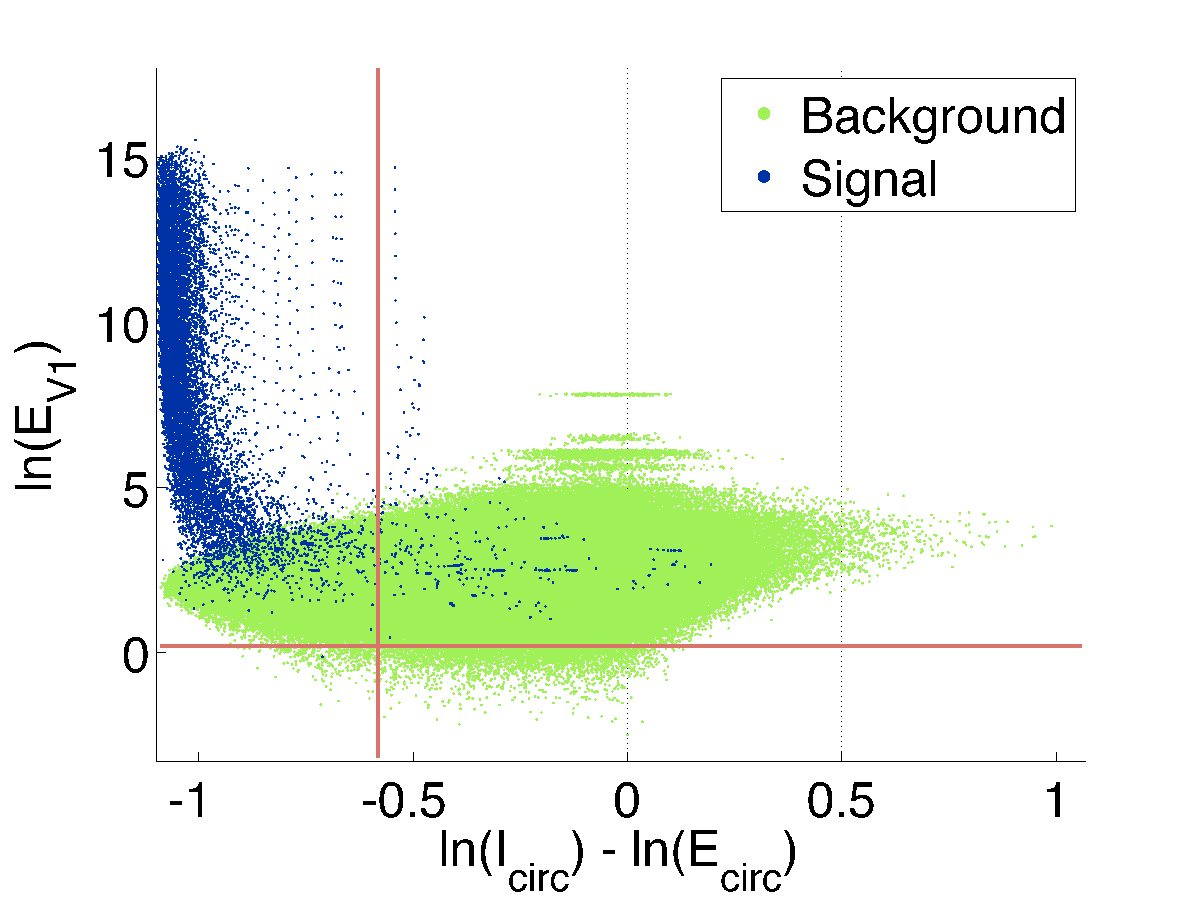}
    \caption{\label{fig:scatter}Scatter plot of signal and background events from the GRB\,060223A analysis.
    The red lines indicate the cuts applied by the decision tree shown in Fig.~\ref{fig:decisiontreeexample1}.}
\end{figure}

\begin{figure*}
    \includegraphics[width=0.95\textwidth]{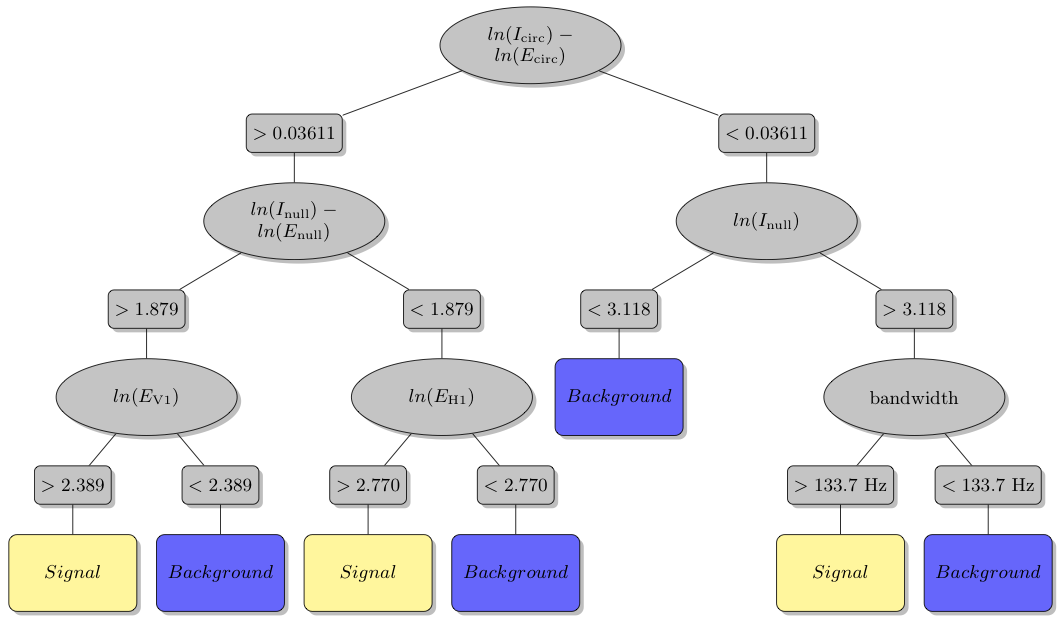}
    \caption{\label{fig:decisiontreeexample2}Example decision tree used in the GRB\,060223A \ac{BDT} analysis.
    Decision nodes are grey ellipses, with thresholds for a branch given in grey rectangles.
    Leaf nodes for signal events are light yellow rectangles and for background events are dark blue rectangles.}
\end{figure*}

Fig.~\ref{fig:decisiontreeexample1} and Fig.~\ref{fig:decisiontreeexample2} show sample decision trees that were used in the GRB\,060223A analysis.
Decision nodes for each variable are shown as grey ellipses with the thresholds for a branch shown in grey rectangles.
The leaf nodes are light yellow rectangles for signal and dark blue rectangles for background.

We first consider the simple decision tree shown in Fig.~\ref{fig:decisiontreeexample1}.
In this example the first cut is applied to the difference between the incoherent and coherent energies, $\ln(I\sub{circ}) - \ln(E\sub{circ})$, at a threshold of $-0.581$.
Events above this threshold are classified as background, while events below this threshold are then cut on the energy in the Virgo detector, $\ln(E\sub{V1})$, which has a threshold of $0.183$.
Events above this second threshold are classified as signal, while events below are classified as background.
The logic behind these choices can be understood from Fig.~\ref{fig:scatter}, which shows a scatter plot of $\ln(E\sub{V1})$ vs.~$\ln(I\sub{circ}) - \ln(E\sub{circ})$ for the training events.
The red lines are the thresholds used in the example decision tree to separate the signal events from the majority of the background events.
While this single tree assigns a large fraction of the background events to the signal leaf node (the upper-left rectangle), the final significance of events is determined collectively by 400 such trees, each generated from a random subset of the training data.
A more complicated tree is shown in Fig.~\ref{fig:decisiontreeexample2}, which classifies events based on 6 of their properties.

\begin{figure}
  \includegraphics[width=0.5\textwidth]{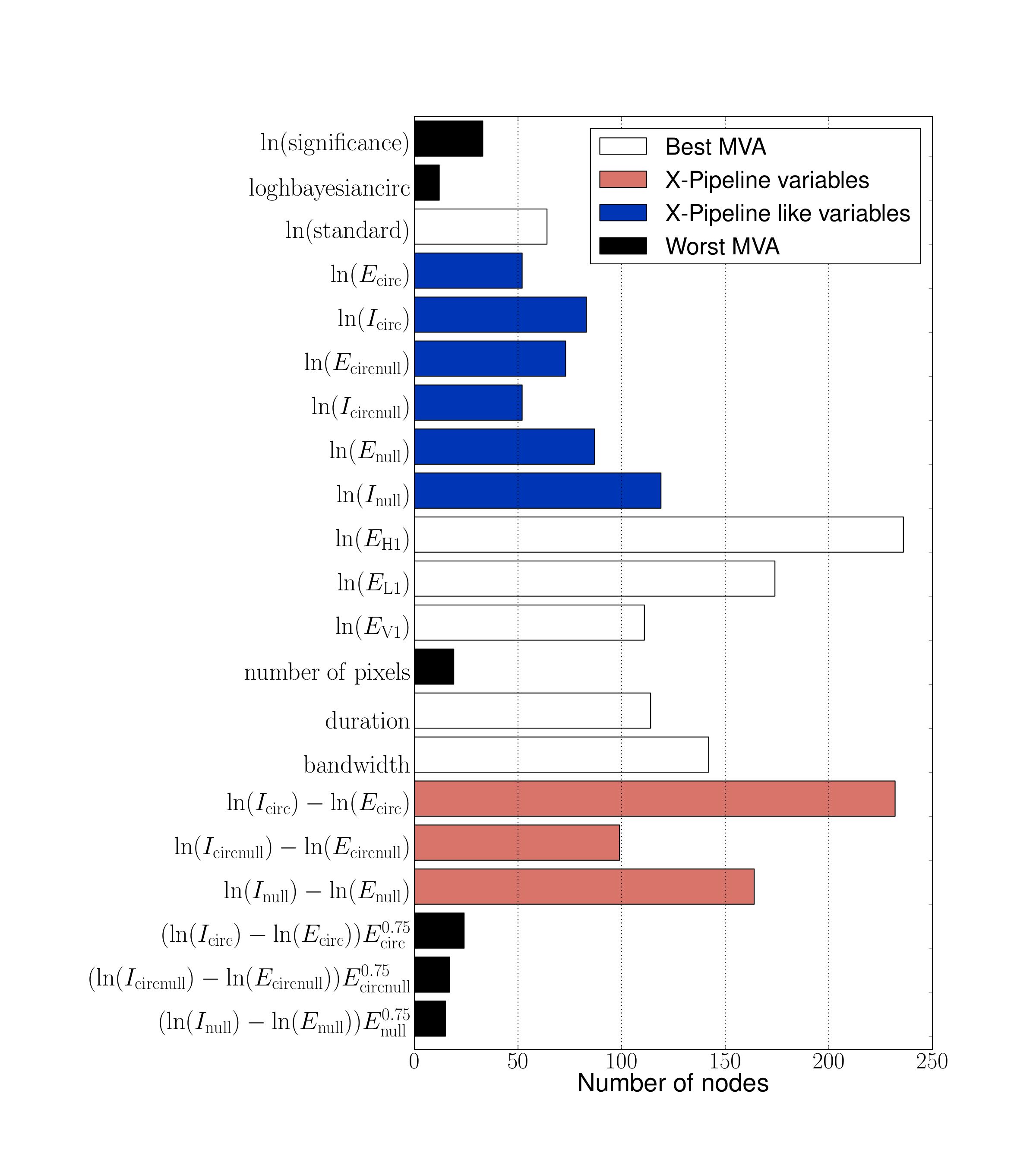}
  \caption{\label{fig:histogram}Bar chart of the number of times each variable or variable combination is used at decision tree nodes in the \ac{BDT} forest for the GRB\,060223A analysis.
The ``\textsc{X-Pipeline} variables'' are combinations of variables that precisely match those used used in the\textsc{X-Pipeline} background rejection test of equation (\ref{eqn:ratiocut}), while the ``\textsc{X-Pipeline} like variables'' can be used to construct the cut of equation (\ref{eqn:alphacut}).  Together these are among the variables most frequently selected by the \ac{BDT} analysis, accounting for half of all nodes.
The single-detector energies, duration, and bandwidth are selected for 40\% of all nodes.  These variables are not used by \textsc{X-Pipeline}
demonstrating how \ac{MVA} makes use of the full dimensionality of the data.
The remaining variables collectively account for approximately 10\% of all nodes.
See Tab.~\ref{tab:properties} for definitions of the variables.}
\end{figure}

We gain further insight into the performance of the \ac{BDT} analysis by considering how frequently different event variables are used in the classification.
Fig.~\ref{fig:histogram} is a bar chart of the total number of times each variable or combination of variables reported by \textsc{X-Pipeline} is used in one of the decision nodes for the GRB\,060223A analysis.
We take this as an indicator of the value of each variable for signal/background discrimination.
The \textsc{X-Pipeline} background rejection tests are based on combinations of $E_\mathrm{circ}$ and $I_\mathrm{circ}$, $E_\mathrm{circnull}$ and $I_\mathrm{circnull}$, and $E_\mathrm{null}$ and $I_\mathrm{null}$, as given in equations (\ref{eqn:ratiocut}) and (\ref{eqn:alphacut}).
The pairwise differences $\ln(I)-\ln(E) = \ln(I/E)$ are labeled as ``\textsc{X-Pipeline} variables'' in the chart, because thresholding on these differences is equivalent to applying the \textsc{X-Pipeline} test of equation (\ref{eqn:ratiocut}).
We see that these are some of the most frequently used combinations, selected for approximately 26\% of all nodes.
The individual $\ln(I)$ and $\ln(E)$ variables are labeled as ``\textsc{X-Pipeline} like variables'' because the remaining \textsc{X-Pipeline} cut of equation (\ref{eqn:alphacut}) can be constructed from them.  These are selected for a total of 24\% of the nodes.
The selection of these variables by \ac{BDT} for approximately 50\% of the nodes affirms their usefulness for signal/background discrimination, as expected from their demonstrated value in \textsc{X-Pipeline}'s background rejection tests.
However, close to half of the \ac{BDT} nodes use variables that are \textit{not} used by \textsc{X-Pipeline}.  This is a clear demonstration of an \ac{MVA} making use of the full dimensionality of the data.
In particular, the single-detector energies are selected for approximately 27\% of all nodes.  Thresholding on these values is equivalent to thresholding on the event signal-to-noise ratio in the individual detectors \cite{Sutton10}, which is not done in \textsc{X-Pipeline}.
The event duration and the bandwidth are also useful, selected for 13\% of all nodes.
The remaining variables collectively account for approximately 10\% of all nodes.
Interestingly, the number of pixels (time-frequency area of the event) is one of the variables that is not particularly useful for signal/background discrimination.

\begin{figure*}
  \begin{center}
    \subfigure[ X-Pipeline]{
      \includegraphics[width=0.45\textwidth]{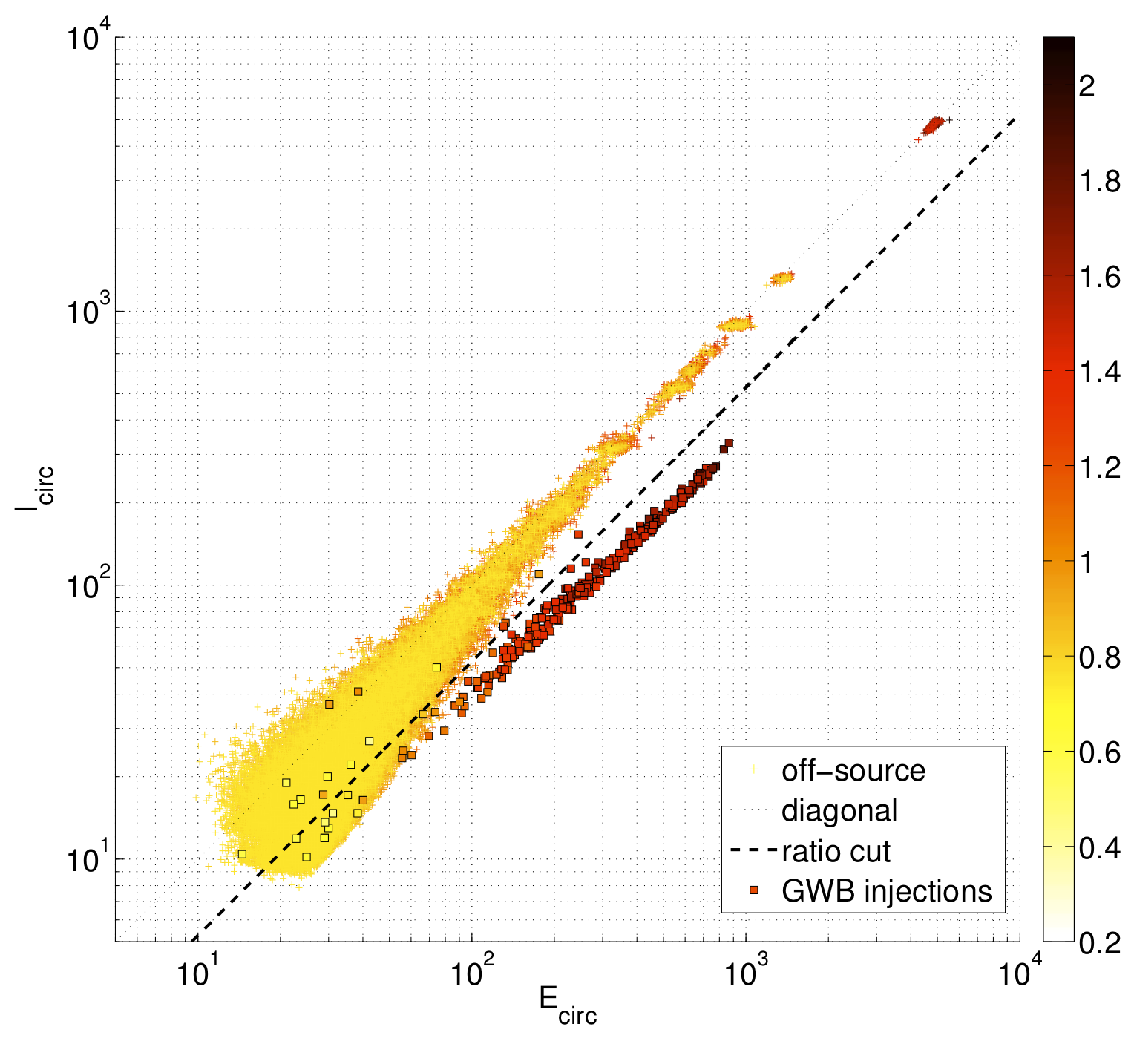}
      \label{fig:Xpipe-offsource}
    }
    \subfigure[ BDT]{
      \includegraphics[width=0.45\textwidth]{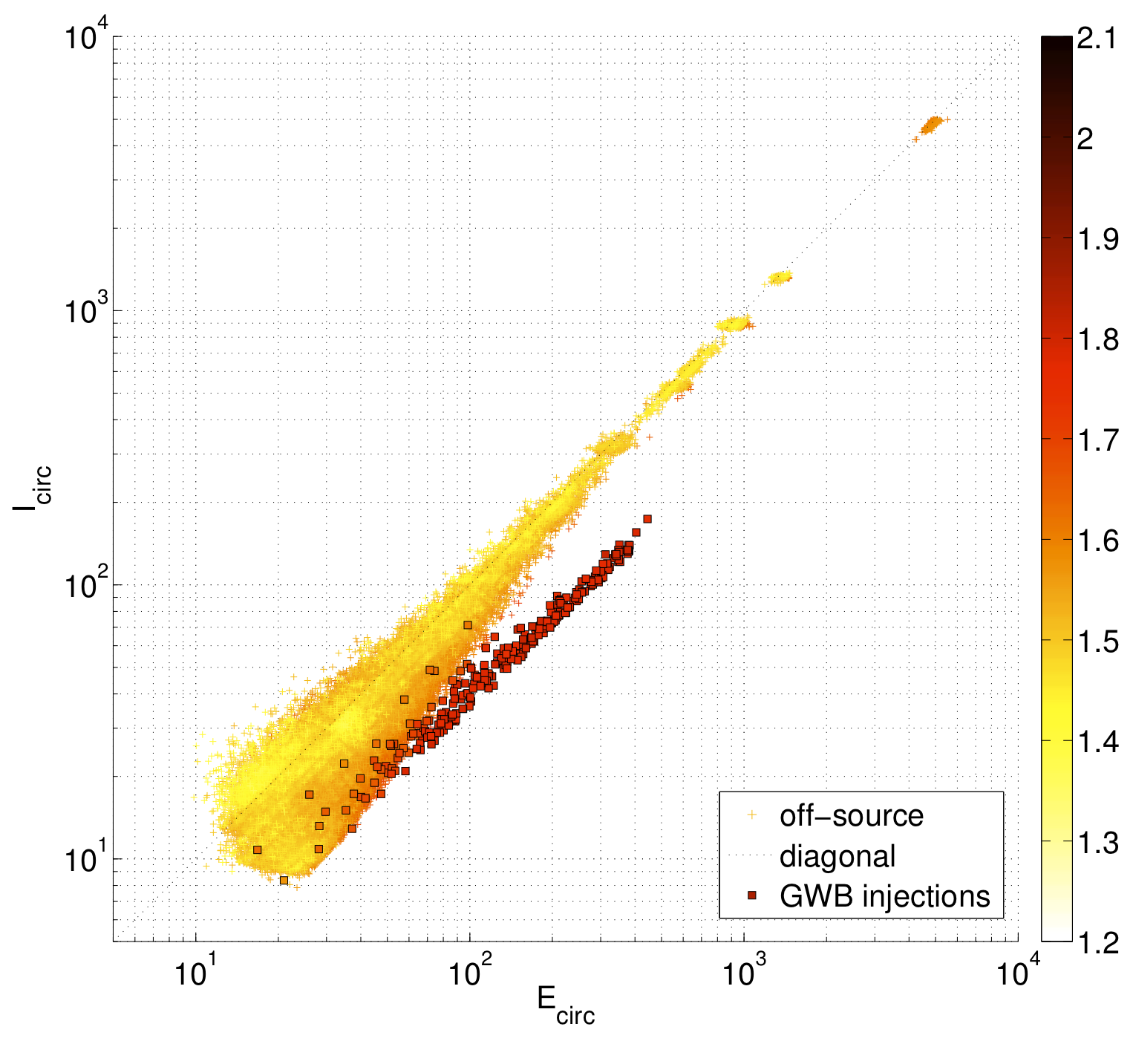}
      \label{fig:MVA-offsource}
    }
  \caption[LoF entry]{\label{fig:offsource}
    (a) - Scatter plot of $I_\mathrm{circ}$ vs.~$E_\mathrm{circ}$ for testing data from the GRB\,060223A \textsc{X-Pipeline} analysis.
    The squares represent simulated \ac{CSG} events at the amplitude for which the \textsc{X-Pipeline} detection efficiency is approximately 90\%.
    Events are coloured by their significance in the \textsc{X-Pipeline} analysis.
    The dashed line is the threshold line for the \textsc{X-Pipeline} background rejection test; events above this line are discarded.
    (b) - Scatter plot of $I_\mathrm{circ}$ vs.~$E_\mathrm{circ}$ for testing data from the GRB\,060223A \ac{BDT} analysis.
    The squares represent simulated \ac{CSG} events at the amplitude for which the \ac{BDT} detection efficiency is approximately 90\%.
    Events are coloured by their significance in the \ac{BDT} analysis.
        }
  \end{center}
\end{figure*}

Another view of the merits of \ac{BDT} classification is given by Fig.~\ref{fig:offsource}.
These show scatter plots of $I_\mathrm{circ}$ vs.~$E_\mathrm{circ}$ for testing data from the GRB\,060223A analysis.
The squares represent simulated \ac{CSG} events at the amplitude for which the detection efficiency is approximately 90\%.
Fig.~\ref{fig:Xpipe-offsource} shows the events coloured by their significance in the \textsc{X-Pipeline} analysis, as well as the threshold line for the \textsc{X-Pipeline} background rejection test.
Event significance increases with $I_\mathrm{circ}$ or $E_\mathrm{circ}$; i.e., along the diagonal, whereas the signal and background are separated primarily in the orthogonal direction.
Signals become detectable when they do not overlap the background distribution in $I_\mathrm{circ}$ vs.~$E_\mathrm{circ}$ space.
Fig.~\ref{fig:MVA-offsource} shows the same background events as ranked by \ac{BDT}.
By contrast with the \textsc{X-Pipeline} analysis, significance increases with distance from the diagonal, so that no additional background rejection test is required.
Simulated \ac{CSG} events are detectable at lower amplitudes, even though they overlap the background distribution in $I_\mathrm{circ}$ vs.~$E_\mathrm{circ}$ space, because the \ac{BDT} analysis takes account of other event properties such as the single-detector energies.

The results presented here indicate that multivariate analysis techniques may be valuable for improving the sensitivity of searches for unmodelled gravitational-wave bursts.
Additional studies are merited, particularly using a wider range of waveform morphologies, larger background samples and lower false alarm rates, and extending to all-sky untriggered searches.
We will revisit these questions in future publications.

\newpage
\section*{Acknowledgments}

The authors would like to thank Michal Was for useful discussions.
We thank the LIGO Scientific Collaboration for permission to use LIGO data for our tests, and for access to the computer clusters on which the analysis was performed.
This work was supported in part by STFC grants PP/F001096/1 and ST/I000887/1.
AM was supported by the NSF's IREU program administered by the University of Florida.

\twocolumngrid
\newpage
\bibliographystyle{unsrt}
\bibliography{references.bib}

\end{document}